\documentclass[aps,prd,twocolumn,reprint]{revtex4-2}

\usepackage{graphicx}% Include figure files
\usepackage{bm}% bold math
\usepackage{hyperref}
\usepackage{amsfonts}
\usepackage{tikz}
\usepackage{amssymb}
\usepackage{graphicx}
\usepackage{color}
\usepackage{epsfig}
\usepackage{remreset}
\usepackage{amsmath,amsthm,ulem}
\usepackage{mathtools}
\usepackage{mathrsfs}
\usepackage{color}
\usepackage{float}
\usepackage{tensor}
\usepackage{lipsum}
\usepackage[caption=false]{subfig}
\usepackage{slashed}
\usepackage{lmodern}
\usepackage[T1]{fontenc}
\usepackage{pgf,pgfarrows,pgfnodes}

\begin{document}

%\preprint{APS/123-QED}

\title{Particle motions and gravitational waveforms in rotating black hole spacetimes of loop quantum gravity}% Force line breaks with \\

\author{Yang Yang}
\email{yangcmy@hotmail.com}
\affiliation{School of Science, Jiangsu University of Science and Technology, 212100 Zhenjiang, China}

\author{Yu-Xuan Bai}
\email{Baiyuxuanzs@hotmail.com}
\affiliation{School of Science, Jiangsu University of Science and Technology, 212100 Zhenjiang, China}

\author{Yong-Zhuang Li}
\email{liyongzhuang@just.edu.cn}
\affiliation{School of Science, Jiangsu University of Science and Technology, 212100 Zhenjiang, China}

\author{Yu Han}%
 \email{hanyu@xynu.edu.cn}
\affiliation{College of Physics and Electrical Engineering, Xinyang Normal University, 464000 Xinyang, China}%

\date{\today}% It is always \today, today,

\begin{abstract}
We study the influence of the loop quantum gravity (LQG) holonomy-correction parameter $\xi$ on black hole horizon structure, timelike geodesic motion, and gravitational wave emission in two rotating LQG-inspired black hole spacetimes, constructed via Newman–Janis algorithm from two distinct spherically symmetric seed metrics (type BH-I and BH-II). The physically admissible range of $\xi$ is determined by requiring the existence of event horizons, marginally bound orbits, and innermost stable circular orbits simultaneously, and is found to shrink monotonically with increasing spin parameter $a$. For equatorial periodic orbits, increasing $\xi$ at fixed angular momentum enlarges the bound energy range, while for off-equatorial orbits, it suppresses the allowed range of the Carter constant, effectively confining trajectories closer to the equatorial plane. The effects of $\xi$ and $a$ on orbital dynamics are systematically antagonistic. Gravitational waveforms computed within a leading-order post-Newtonian extreme-mass-ratio inspiral (EMRI) model show that larger $\xi$ produces enhanced deviations from the Kerr waveform, more prominently so for type BH-II than type BH-I. The resulting characteristic strains occupy the $(10^{-3}, 0.1)$ Hz frequency band but fall below the sensitivity curves of current and near-future space-based detectors for the EMRI parameters considered ($M = 10^7 M_\odot$, $m = 10 M_\odot$, $D_L = 200$ Mpc). Adiabatic inspiral calculations confirm that $\xi$ and $a$ drive orbital evolution in opposite directions, with their relative magnitude determining whether quantum corrections accelerate or retard the inspiral. These results establish systematic observational signatures of holonomy corrections in rotating LQG black holes and motivate higher-fidelity waveform modeling for future space-based gravitational wave detectors.
\end{abstract}

\maketitle

%\tableofcontents

\section{\label{sec:intro}Introduction}

The detection of gravitational waves (GWs) by the LIGO and Virgo collaborations has ushered in a new era of observational astrophysics, providing unprecedented access to the dynamics of strong-field gravity and the nature of compact objects such as black holes or neutron stars \cite{LIGOScientific:2016aoc,LIGOScientific:2016sjg,LIGOScientific:2016lio}. In particular, the observation of binary black hole mergers and extreme mass-ratio inspirals (EMRIs) have opened a unique window into the geometry of spacetime near event horizons. These systems serve as ideal laboratories for testing general relativity (GR) or modified gravity theories in its most extreme regime and for probing potential quantum gravitational corrections to classical spacetime structure \cite{Will:1994fb,Saijo:1996iz,Yunes:2013dva,Berti:2015itd,LIGOScientific:2019fpa,Brahma:2020eos,LIGOScientific:2020tif,Agullo:2020hxe,Krishnendu:2021fga,LIGOScientific:2021sio,Li:2024tld,Yang:2024lmj,Li:2025sfe,Chen:2025aqh,Ahmed:2025shr,Deppe:2025pvd}.

Among the leading candidates for a quantum theory of gravity, loop quantum gravity (LQG) is the main representative of the background-independent, non-perturbative approach that quantizes spacetime geometry itself, replacing the smooth continuum of GR with a discrete, spin-network-based structure at the Planck scale \cite{Perez:2004hj,Ashtekar:2011ni,Ashtekar:2021kfp}.  The classical singularity at the center of a black hole can be resolved through quantum geometry effects in LQG, leading to the emergence of a “quantum bounce” and a transition to an expanding region, thereby replacing the singularity with a bridge to another spacetime or a “white hole”-like region \cite{Bojowald:2018xxu,Ashtekar:2018lag,Alesci:2019pbs,Assanioussi:2019twp,Han:2023wxg}. Furthermore, LQG predicts modifications to the event horizon structure and the near-horizon geometry of black holes, which manifest as deviations from the GR or other classical modified gravity theories. These deviations, while negligible at distance far from the horizon, are expected to become dominant in the strong-field regime, potentially leaving imprints on the gravitational waveforms emitted by orbiting test particles.

From an astrophysical perspective, black holes are generally believed to possess spin characteristics. However, even static spherically symmetric non-rotating black hole models have been extensively discussed in the context of LQG \cite{Modesto:2005zm,Perez:2017cmj,Gan:2020dkb,Borges:2023fub,Zhang:2023yps}, the construction of axisymmetric spacetimes within this framework remains one of the most challenging open questions, for the reasons such as the inherent complexity of incorporating angular momentum into the discrete quantum geometry framework where rotational symmetry is broken at the quantum level due to the underlying spin network structure \cite{Frodden:2012en,Gambini:2018ucf,Gambini:2020fnd}. A fully self-consistent, quantum-gravity-derived rotating black hole solution that is both analytically tractable and physically well motivated remains elusive. Previous attempts at generating rotating spacetimes in LQG are mainly focusing on the so-called the Newman-Janis algorithm (NJA) \cite{Liu:2020ola,Jiang:2023img,Li:2024ctu,Ali:2024ssf,Mustafa:2025mkc,Sekhmani:2025bsi}. Although this NJA may cause certain physical issues in specific contexts, its modified versions can still serve as a useful tool for constructing plausible effective models that explore potential quantum gravity corrections deviating from the Kerr black hole paradigm \cite{Hansen:2013owa,Azreg-Ainou:2014pra,Erbin:2016lzq,Junior:2020lya,Chaturvedi:2023ctn}.

Aiming to bridge theoretical frameworks with astrophysical data, numerous investigations have been undertaken to explore the observational and phenomenological effects of LQG-inspired black holes, such as those linked to quasi-normal modes \cite{Fu:2023drp,Moreira:2023cxy,Bolokhov:2023bwm,Yang:2024ofe,Gingrich:2024tuf}, gravitational lensing \cite{Liu:2024wal,Sahu:2015dea,Fu:2021fxn,Vagnozzi:2022moj,Afrin:2022ztr,Huang:2022iwl,KumarWalia:2022ddq,Junior:2023xgl,Kumar:2023jgh,Soares:2024rhp,Dong:2024alq,Zhao:2024elr,Ahmed:2024fye,Calza:2024xdh,Jiang:2024cpe,Li:2024afr,Alloqulov:2025htt,Mushtaq:2025shw,Al-Badawi:2025yum,Huang:2025gia,Soares:2025hpy,Wang:2025fmz} and the test particle motions \cite{Ahmed:2025shr,Alloqulov:2025bxh,Guo:2025scs,Chen:2025baz,Huang:2026igb,Agrawal:2026rwu,Chen:2026kbn}. In this article, we investigate the particle motions and the resulting gravitational waveforms in two different  NJA-constructed LQG-inspired rotating black hole (LQGBH) spacetimes, in which the quantum effects (quantified by a single regularization  parameter) rapidly die out when moving away from the center, with a well-defined asymptotic region in the exterior \cite{Brahma:2020eos}. Meanwhile, the generated LQG-inspired rotating black holes have been proved to be more general and captured some universal properties of rotating LQGBHs. As aforementioned, the gravitational waveforms emitted from EMRIs encode rich information about the features of the central
objects, and inherit distinctive imprints of periodic orbits which are bound trajectories of test particles returning to its initial state after completing an integer number of radial and angular oscillations. Such investigations provide a potential observational pathway to test LQG predictions with future space-based gravitational wave detectors (such as  Taiji, TianQin, and LISA) and deepens our understanding of dynamics in quantum-modified spacetimes \cite{TianQin:2015yph,Hu:2017mde,Babak:2017tow,Hughes:2000ssa,Gong:2021gvw,Torres-Orjuela:2023hfd}.

The paper is organized as follows. In Sec. \ref{sec:kerrlikes} we briefly review the properties of the two different rotating LQG-inspired black holes constructed via Newman-Janis algorithm, then we investigate the influences of the parameter $\xi$ on the timelike geodesics of the particles in Sec. \ref{sec:dynamassive}, mainly focusing on the prograde periodic orbits. Then, we examine the gravitational waveforms that radiate from the prograde periodic bound orbits in one complete period of a test object in Sec. \ref{sec:GW}, and investigate the evolution of the apastron with the adiabatic approximation in Sec. \ref{sec:evop}. Finally we give summaries and remarks in Sec. \ref{sec:con}. Unless otherwise specified, all quantities are re-scaled in units of $M$ with $M=1$ in the numerical calculations.

\section{The LQG-inspired rotating black holes}\label{sec:kerrlikes}

The seed metrics describing the spherically symmetric black holes we adopt in this article are given by  \cite{Zhang:2024khj,Zhang:2024ney,Zhang:2025ccx,Chen:2025aqh}:

\begin{eqnarray}\label{eq:Schmetric}
    ds^{2}=-f(r)dt^2+\frac{dr^2}{g(r)}+r^2(d\theta^2+\sin^{2}{\theta}d\phi^2),
\end{eqnarray}
where for black hole of type I (BH-I),
\begin{eqnarray}
    f(r)=g(r)=1-\frac{2M}{r}+\frac{\xi^2}{r^2}\left(1-\frac{2M}{r}\right)^2,\nonumber\\
\end{eqnarray}
and for black hole of type II (BH-II),
\begin{eqnarray}
    f(r)&=&1-\frac{2M}{r},\nonumber\\
    g(r)&=&1-\frac{2M}{r}+\frac{\xi^2}{r^2}\left(1-\frac{2M}{r}\right)^2,
\end{eqnarray}
In which $\xi$ denotes the regularization parameter in the holonomy correction, i.e.  the replacement of the extrinsic $k$ by its holonomy
\begin{eqnarray}
k\rightarrow\frac{r}{\xi}\sin{\frac{\xi k}{r}}.
\end{eqnarray}
It should be noted that although in the full theory of LQG the holonomy corrections arise from the discrete spin-network structure so $\xi$ should be fixed by the spin network scale, the relations between the full theory and the symmetry reduced models such as the black holes are still not sufficiently studied, therefore, in the current research status of LQGBH, there is still no compelling theoretical constraint on the value of $\xi$. 

Following procedure presented in Ref. \cite{Brahma:2020eos}, the metric of the revised Newman–Janis algorithm (NJA)-induced rotating LQGBH is given by
\begin{eqnarray}\label{eq:rmetric}
    ds^{2}=&-&\left(1-\frac{2 M(r) r}{\rho^{2}}\right)dt^{2}-\frac{4aM(r)r\sin^{2}{\theta}}{\rho^2}dtd\phi\nonumber \\
   & +&\rho^2d\theta^{2}+\frac{\rho^{2}dr^2}{\Delta}+\frac{\Sigma \sin^{2}{\theta}}{\rho^2}d\phi^2,
\end{eqnarray}
where
\begin{eqnarray}
    \rho^{2}&=&K+a^2 \cos^{2}{\theta},\,\quad M(r)=\frac{K-g(r)r^{2}}{2r},\nonumber \\
    \Delta&=&g(r)r^{2}+a^2,\,\quad \Sigma=(K+a^2)^{2}-a^{2}\Delta \sin^{2}{\theta},
\end{eqnarray}
where $K\equiv r^2\sqrt{g/f}$ and the spin parameter $a\equiv \frac{J}{M}$ with $J$ the angular momentum of the black hole.
In the case $\xi\rightarrow 0$, one has $f(r)=g(r)=1-2M/r$ and  the metric (\ref{eq:rmetric}) reduces to  the classical Kerr black hole metric. However, with the spin approaches to zero, i.e., $a\rightarrow0$, the metric (\ref{eq:rmetric}) reduces to the Schwarzschild one for type BH-I but to a Schwarzschild black hole with a conformal factor $\sqrt{g(r)/f(r)}$ for type BH-II, i.e., $ds^2=\sqrt{g(r)/f(r)}ds^{2}_{Sch}$ where $ds^{2}_{Sch}$ is the metric (\ref{eq:Schmetric}) for BH-II. Given that we are considering the case of axisymmetric spacetimes, we do not multiply the metric (\ref{eq:rmetric}) by the conformal factor $\sqrt{f(r)/g(r)}$ to ensure that it reduces to the classical Kerr black hole in the limit $\xi\rightarrow0$.

Furthermore, the Arnowitt-Deser-Misner (ADM) mass of the rotating LQGBH can be calculated as \cite{Wald:1984rg}
\begin{eqnarray}
    M_{ADM}=\lim_{r\rightarrow\infty}\frac{r}{2}\left(g(r)^{-1}-1\right)=M,
\end{eqnarray}
i.e., the ADM mass of the rotating LQGBH equals the black hole mass which is a Dirac observable. 

The definition of $K$ requires  $g/f>0$, which should be regarded as a condition that constrains the range of $r$ in which the a well-defined metric must satisfy. For BH-I, this condition is automatically satisfied for arbitrary $r$.  For BH-II, this condition leads to a restriction, i.e., $\xi^2 (r-2M)+r^3>0$, which is naturally satisfied for $r>2 M$, but for $r<2 M$ we obtain the range  $r^3/(2M-r)>\xi^2$ for a well defined metric. 

If we assume that a realistic rotating LQGBH has an event, it will suggest a restriction condition for $\xi$ and $a$. From the metric (\ref{eq:rmetric}), the radius of the horizon is determined by $\Delta=0$, which corresponds to finding the roots of the equation
\begin{eqnarray}\label{eq:souhorizon}
    h(r)&=&r^4-2Mr^3+r^2(\xi^2+a^2)-4Mr\xi^2+4M^{2}\xi^2 \nonumber \\
    &=&0.
\end{eqnarray}

From the argument in Appendix A, it is not difficult to see that in order to have an event horizon we must require $a<M$.
Since in loop quantum gravity the regularization parameter $\xi$ should be real, Eq. (\ref{eq:souhorizon}) unequivocally indicates that $h(r)>0$ and $h'(r)<0$ as $r\rightarrow0$. Moreover, given that $h(r)\rightarrow\infty$ and $h'(r)\rightarrow\infty$ as $r\rightarrow\infty$, the function $h(r)$ must possess at least one local minimum. 
%\begin{figure}
%    \centering
%    \includegraphics[width=\linewidth]{Fig-horizon.pdf}
%    \includegraphics[width=\linewidth]{Fig-horizon2.pdf}
%    \caption{The upper panel shows the shape of function $h(r)$ for different $\xi$ %with $a/M=0.9$, while the lower panel shows the shape of function $h(r)$ for different %$a$ with $\xi/M=0.8$. The intersection points of $h(r)$ with the x-axis denote the %positions of the Cauchy horizon and the event horizon respectively. }
%    \label{fig:horizon}
%\end{figure}
In the case of the extremal black hole, We denote the horizon radiusby $r_c$ and the corresponding parameter $\xi$ by $\xi_e$.
To find  all possible extreme points $(r_c,\,\xi_e)$ for given $a/M$ , we  consider a special condition that the following equations
\begin{eqnarray}
h(r_c)=h'(r_c)=0
\end{eqnarray}
are simultaneously satisfied. In fact, it is shown in  Appendix A that for given $a$ and $M$  only a single set of roots $(r_c,\,\xi_e)$ exists, which means that in this case the rotating LQGBH has only one horizon, with the extremal horizon radius $r_c$  given by
\begin{eqnarray}\label{eq:criticalr2}
    r_c=\frac{M}{3}\left(5-2\sqrt{7} \cos{\frac{\pi+\delta}{3}}\right),
\end{eqnarray}
in which 
\begin{eqnarray}
\delta=\arccos{\frac{27a^2-10M^2}{7\sqrt{7}M^2}},  
\end{eqnarray}
with $a<M$, and the corresponding value of $\xi_{e}$ is given by
\begin{eqnarray}\label{eq:xisqc}
 \xi^2_e=\frac{r^2_c[2r_cM-r^2_c-a^2)]}{(r_c-2M)^2}.
\end{eqnarray}
or, equivalently,
\begin{eqnarray} \label{eq:xisqc2}
 \xi^2_e=-\frac{r^3_c(r_c-M)}{2M(r_c-2M)}.
\end{eqnarray}
From Eqs. (\ref{eq:xisqc}) and (\ref{eq:xisqc2}), we find that when $a/M\rightarrow 1$, we have 
\begin{eqnarray} \label{eq:apprrc}
\xi_e^2 \simeq r^2_c-a^2.
\end{eqnarray}

\begin{figure}
    \centering
    \includegraphics[width=\linewidth]{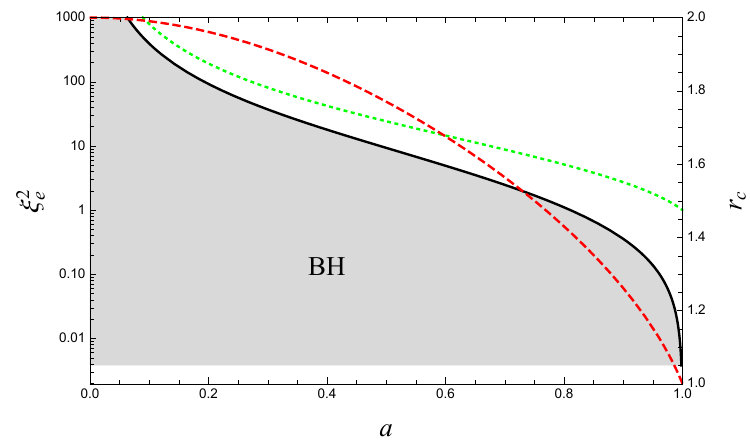}
    \caption{The critical values of $r_c$ and $\xi^{2}_{e}$  as functions of $a$ for given $M$ (which is set to $1$ unless specified). The red dashed line (corresponding to the right axis) shows the critical $r_c$ as a function of $a$. The black line (corresponding to the left axis) shows the critical $\xi^{2}_{e}$ as a function of $a$, for each value of $\xi$ in the gray region the rotating LQGBH has an event horizon and a Cauchy horizon.  The region below the green dotted line shows the restriction condition given by $\xi^2<-r_c^3/(r_c-2M)$ for a well-defined metric of BH-II.}
    \label{fig:criticalxi}
\end{figure}

In Fig. \ref{fig:criticalxi}, we plot the values of $\xi^2_{e}$ and $r_c$ as a function of $a$ for given $M$ , from which we observe that the requirement of at least one Cauchy horizon and one event horizon in the LQG black hole requires $\xi$ to decrease monotonically with increasing $a$, as $a/M$ approaches zero, $\xi$ may span the entire real axis; as $a/M\rightarrow1$,  $\xi$ approaches zero. Furthermore, we also observe that the extremal horizon radius also decreases with increasing $a$.

It should be noted that the above restriction on $\xi$ is constructed through the metric solely. In the following, Further restrictions will be provided by other conditions.

\section{PRECESSING AND PERIODIC ORBITS OF THE TEST PARTICLES}\label{sec:dynamassive}
In this section, we investigate the geodesic motion for a test particle in the rotating LQGBH spacetime. The Lagrangian  can be constructed from the metric (\ref{eq:rmetric}) as
\begin{eqnarray}\label{eq:Lagrangian}
    2\mathcal{L}=&-&\left(1-\frac{2 M(r) r}{\rho^{2}}\right)\dot{t}^{2}-\frac{4aM(r)r\sin^{2}{\theta}}{\rho^2}\dot{t}\dot{\phi}\nonumber \\
   & +&\rho^2\dot{\theta}^{2}+\frac{\rho^{2}\dot{r}^2}{\Delta}+\frac{\Sigma \sin^{2}{\theta}}{\rho^2}\dot{\phi}^2,
\end{eqnarray}
where the dot denotes the derivative with respect to the affine parameter $\lambda$. For a more convenient derivation of the geodesic equations, three constants of motion are traditionally required. The first two constants,i.e.  the energy $E$ and the angular momentum $L$ per unit mass of the particle, originate from the two killing vectors related to the time and azimuth angle translation invariance of the spacetime,
\begin{eqnarray}
    \frac{\partial\mathcal{L}}{\partial\dot{t}}&=&g_{tt}\dot{t}+g_{t\phi}\dot{\phi}\equiv-E,\label{eq:conservedq1}\\
    \frac{\partial\mathcal{L}}{\partial\dot{\phi}}&=&g_{\phi\phi}\dot{\phi}+g_{t\phi}\dot{t}\equiv L.\label{eq:conservedq2}
\end{eqnarray}
The third constant,  namely the Carter constant \cite{Carter:1968rr}, originates from the separability of the Hamilton–Jacobi equation. To observe this, we take the ansatz that the  Hamilton–Jacobi action can be written as
\begin{eqnarray}
    S=-\frac{1}{2}m^{2}\lambda-Et+L\phi+S_{r}(r)+S_{\theta}(\theta).
\end{eqnarray}
Then, from the Hamilton–Jacobi equation we have
\begin{equation}
    \frac{\partial S}{\partial \lambda}=-\frac{1}{2}g^{\mu\nu}\frac{\partial S}{\partial x^{\mu}}\frac{\partial S}{\partial x^{\nu}}.
\end{equation}
Substituting the metric (\ref{eq:rmetric}) into the above  and separating the variables yields the following equation:
\begin{eqnarray}
    \left(\frac{dS_{r}}{dr}\right)^{2}&=&\frac{1}{\Delta^{2}}\Big [\big(E (K+a^{2})-aL\big)^{2}-\Delta\big(\mathcal{C}+K\nonumber\\
    && \quad\quad\quad -2aEL\big)\Big ],\\
    \left(\frac{dS_{\theta}}{d\theta}\right)^{2}&=&\mathcal{C}-\left(a^{2}E^{2}\sin^{2}{\theta}+L^{2}\csc^{2}{\theta}+a^2\cos^{2}{\theta}\right),\nonumber\\
\end{eqnarray}
where $\mathcal{C}$ is the Carter constant. It should be noted that the separability of the Hamilton–Jacobi equation is independent of the specific form of $K$, implying that both the type I and type II black hole configurations are integrable in this sense. 

Using the relations
\begin{eqnarray}
    \frac{\partial S}{\partial x^{\mu}}=g_{\mu\nu}\dot{x}^{\nu},
\end{eqnarray}
and Eqs. (\ref{eq:conservedq1}, \ref{eq:conservedq2}), we finally obtain the geodesic equations 
\begin{eqnarray}
    \rho^{2}\dot{t}&=&\frac{1}{\Delta}\left(E\Sigma-2arM(r)L\right),\label{eq:geodesiceq1}\\
    \rho^{2}\dot{\phi}&=&\frac{1}{\Delta}\left[2arM(r)E+(\rho^2-2rM(r))L\csc^{2}{\theta}\right],\label{eq:geodesiceq2}\\
    \rho^{4}\dot{r}^{2}&=&\mathcal{R}(r),\label{eq:geodesiceq3}\\
    \rho^{4}\dot{\theta}^{2}&=&\mathcal{Q}(\theta),\label{eq:geodesiceq4}
\end{eqnarray}
in which
\begin{eqnarray}
    \mathcal{R}(r)&\equiv&\left[E (K+a^{2})-aL\right]^{2}-\Delta\left(\mathcal{C}+K-2aEL\right),\nonumber\\ \label{eq:geodesiceq5}\\
    \mathcal{Q}(\theta)&\equiv&\mathcal{C}-\left(a^{2}E^{2}\sin^{2}{\theta}+L^{2}\csc^{2}{\theta}+a^2\cos^{2}{\theta}\right).\label{eq:geodesiceq6}
\end{eqnarray}

\subsection{The particle motions on the equatorial plane}
For simplicity, we begin with the investigations of the particle motions on the equatorial plane on which $\theta=\pi/2$ and $\dot{\theta}=0$, in this case, the equation of radial motion can be expressed as
\begin{eqnarray}\label{eq:radialeq}
    \dot{r}^2&=&E^{2}-\frac{\Delta K-K(a^2E^2-L^2)-2M(r)r(aE-L)^2}{K^2} \nonumber \\
    &=&E^2-V_{eff}(r),
\end{eqnarray}
where $V_{eff}(r)$ is called the effective potential. Two special bound orbits can be determined through the above equation. The first are the marginally bound orbits (MBOs) which specifically refer to the finally captured unstable circular orbits around the black hole formed by a test particle falling freely from the infinity, satisfying $\dot{r}=0, E=1$;  the second are the innermost stable circular orbits (ISCOs) which are the closest stable circular orbits that a particle can orbit the black hole, satisfying $\dot{r}=0, \ddot{r}=0$.

\begin{figure*}
    \centering
    \includegraphics[width=\textwidth]{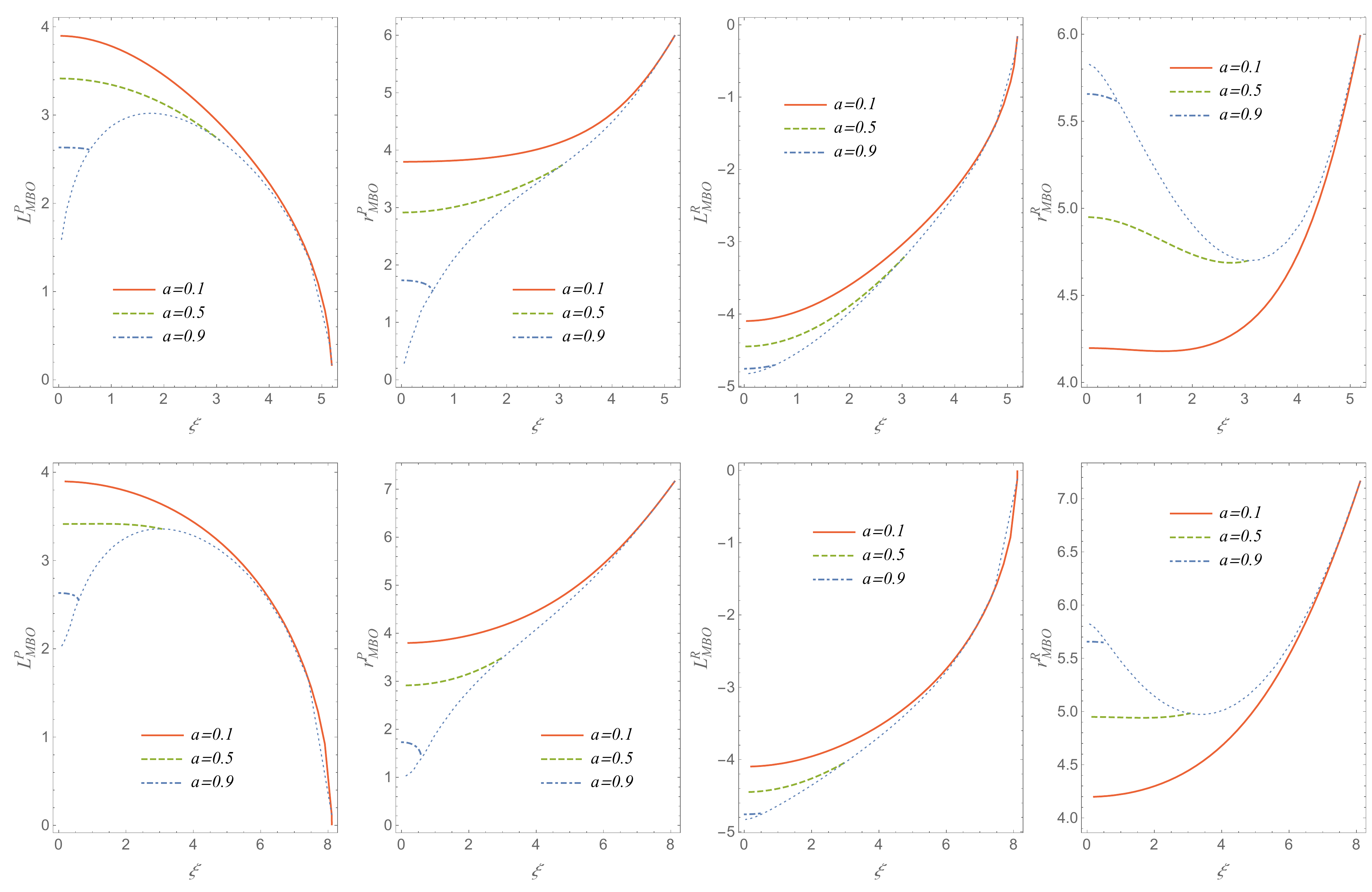}
    \caption{The angular momentum $L_{MBO}$ and the radius $r_{MBO}$ of the particles on the MBOs as functions of $\xi$ with $a=0.1,\,0.5,\,0.9$. The right superscripts $P$ and $R$ represent the prograde and retrograde orbits, respectively. The first row is calculated corresponding to BH-I, and the second row corresponds to BH-II. The dotted blue thin line shows the new restriction on $\xi$, which is constructed under the condition that the MBOs exist with given $a/M$.}
    \label{fig:MBO}
\end{figure*}

For the MBOs, the determining conditions $\dot{r}=0, E=1$ can be re-expressed as 
\begin{eqnarray}\label{eq:MBO}
    V_{eff}(r)=E^{2}=1,\quad V'_{eff}(r)=0,
\end{eqnarray}
where the prime represents the derivative with respect to $r$. 

Note that Eq. (\ref{eq:MBO})  sets a new restriction for $\xi$. Indeed, for BH-I, in the limit $L\rightarrow0$, the only none zero solution of Eq. (\ref{eq:MBO}) is 
\begin{eqnarray}
   r=6 M,\quad \xi=3\sqrt{3}M,
\end{eqnarray}

 Combining with the restrictions (\ref{eq:criticalr2},\ref{eq:xisqc}), the critical $\xi_{c}$ should then be given by $\xi_{c}=\min\left(\xi_e,\,3\sqrt{3}M\right)$. Moreover, for  BH-II, one can obtain $\xi_{c}=\min\left(\xi_e,\, M\sqrt{4(223+70\sqrt{10})/27}\right)$. 
 
 In Fig. \ref{fig:MBO} we show the effects of $\xi$ on the critical angular momentum $L_{MBO}$ and the radius $r_{MBO}$ of the particles on the MBOs with $a=0.1,\, 0.5,\,0.9$, either prograde or retrograde relative to the black hole's spin direction. Interestingly, such effects are quite distinct for different $a$ but are similar between both types of black holes.

Now, we consider the ISCOs, which are determined by
\begin{eqnarray}\label{eq:ISCO}
    V_{eff}(r)=E^{2},\,V_{eff}'(r)=V_{eff}''(r)=0,
\end{eqnarray}
from which we can fix three quantities $r_{ISCO}$, $E_{ISCO}$, $L_{ISCO}$ for the particles moving along the ISCOs.

\begin{figure*}
    \centering
    \includegraphics[width=\textwidth]{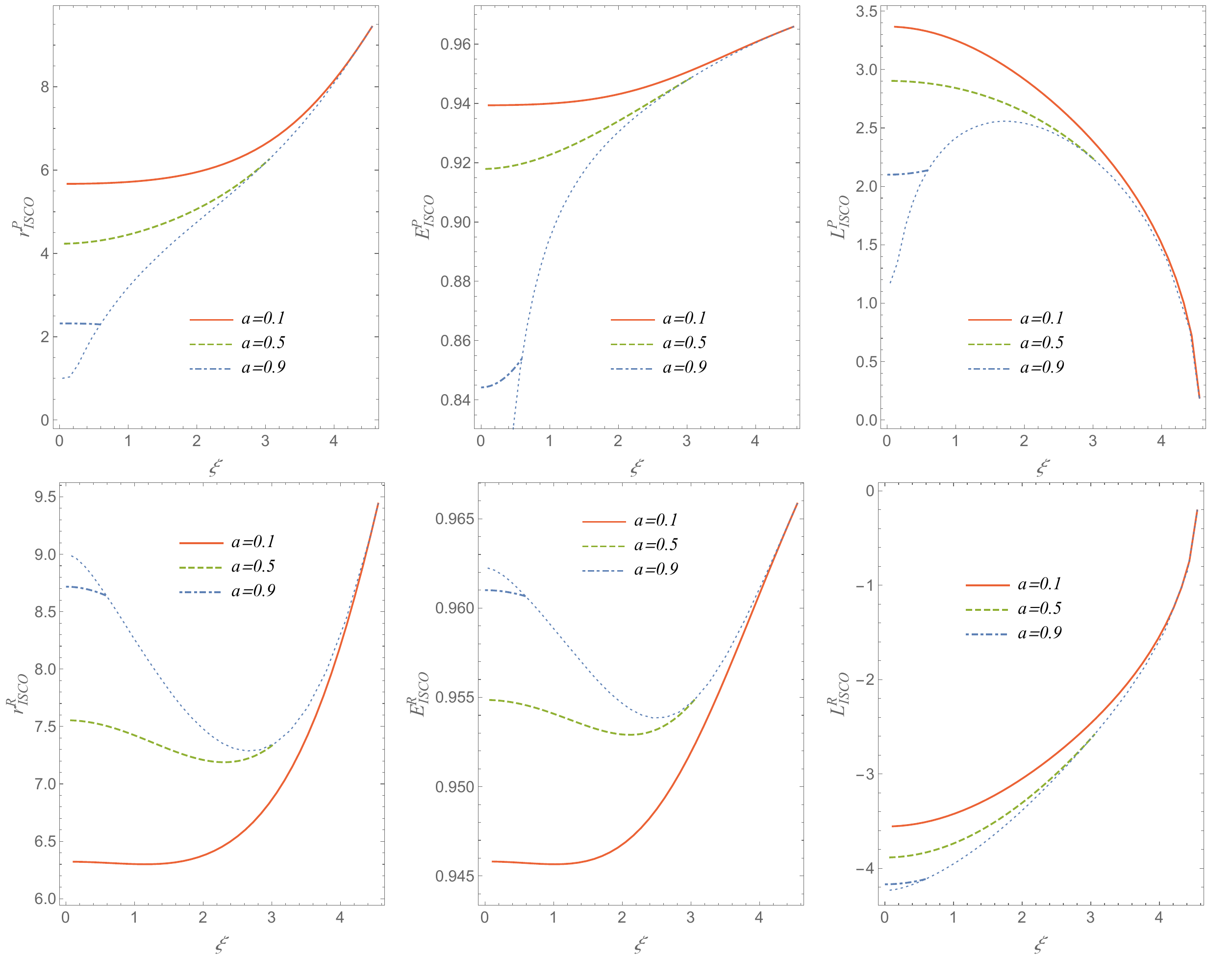}
    \caption{The angular momentum $L_{ISCO}$, the energy $E_{ISCO}$ and the radius $r_{ISCO}$ of the particles on the ISCOs as functions of $\xi$ with $a=0.1,\,0.5,\,0.9$ for BH-I. The right superscripts $P$ and $R$ represent the prograde and retrograde orbits, respectively. The dotted blue thin line shows the new restriction on $\xi$, which is determined under the requirement that the ISCOs exist for arbitrary given $a/M$.}
    \label{fig:ISCO-I}
\end{figure*}

\begin{figure*}
    \centering
    \includegraphics[width=\textwidth]{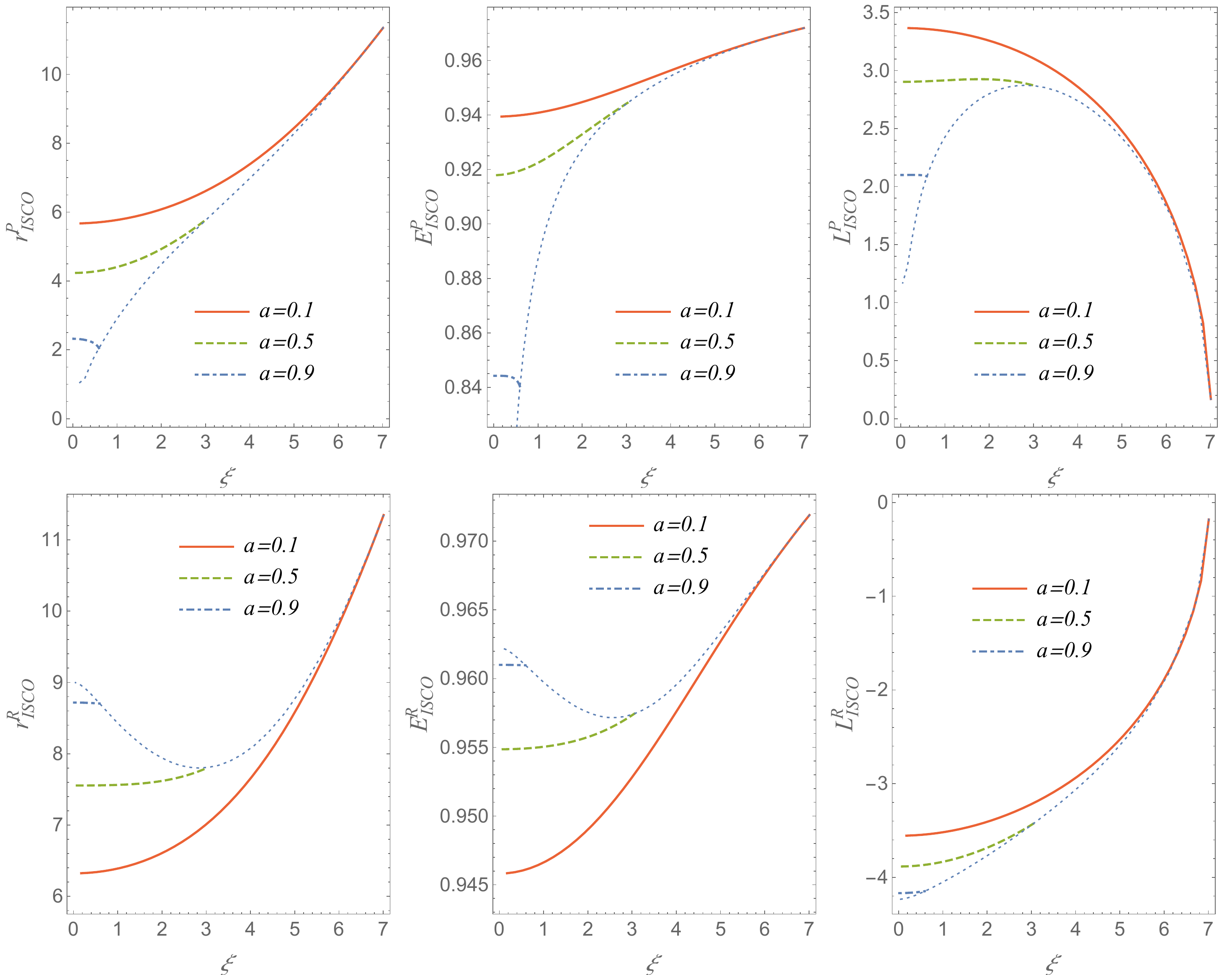}
    \caption{The angular momentum $L_{ISCO}$, the energy $E_{ISCO}$ and the radius $r_{ISCO}$ of the particles on the ISCOs as functions of $\xi$ with $a=0.1,\,0.5,\,0.9$ for BH-II. The right superscripts $P$ and $R$ represent the prograde and retrograde orbits, respectively. The dotted blue thin line shows the new restriction on $\xi$,  which is determined under the requirement that the ISCOs exist for arbitrary given $a/M$.}
    \label{fig:ISCO-II}
\end{figure*}
The dependence of these three quantities on $\xi$ are plotted in Fig. \ref{fig:ISCO-I} for black hole of type I and in Fig. \ref{fig:ISCO-II} for type II. Interestingly, one can observe for both types the behaviors of $r_{ISCO}$ and $E_{ISCO}$ with $\xi$ remain remarkably similar, regardless of whether the particle’s trajectory is prograde or retrograde, but the behavior of $L_{ISCO}$ is quite different from the other two quantities. 

The critical $\xi_{c}$ of ISCOs is given by $\xi_{c}=\min{\left(\xi_e,\,\sqrt{12\sqrt{3}}M\right)}$ for black hole of type I and $\xi_{c}=\min(\xi_e,\, 2\sqrt{28+19\sqrt{19}}M/3)$ for type II. We remind the reader here that such critical values are calculated by requiring the existences of the ISCOs.

In addition to the two aforementioned special types of bound orbits, another distinctive bound orbits, namely the precessing and periodic orbits also exist. For the test particles moving on the equatorial plane, the motions are completely determined by the evolution of $r$ and $\phi$, and can be simply described by an unique number $q$ as \cite{Levin:2008mq} 
\begin{eqnarray}
    q=\frac{\Delta \phi}{2\pi}-1,
\end{eqnarray}
where $\Delta\phi$ is the accumulated azimuth between two turning points, i.e., the periastron $r_p$ and apastron $r_a$ of the bound orbit during a radial period
\begin{eqnarray}
    \Delta\phi=2\int^{r_a}_{r_p}\frac{d\phi}{dr}dr.
\end{eqnarray}
Clearly, if $q$ is a rational number, the orbits will be periodic and return to its initial location exactly after a finite time; but for non-rational $q$, the test particle will run a precessing orbit around the black hole with the precession per revolution $\Delta\chi=\Delta\phi-2\pi$.

\begin{figure*}
    \centering
    \includegraphics[width=\textwidth]{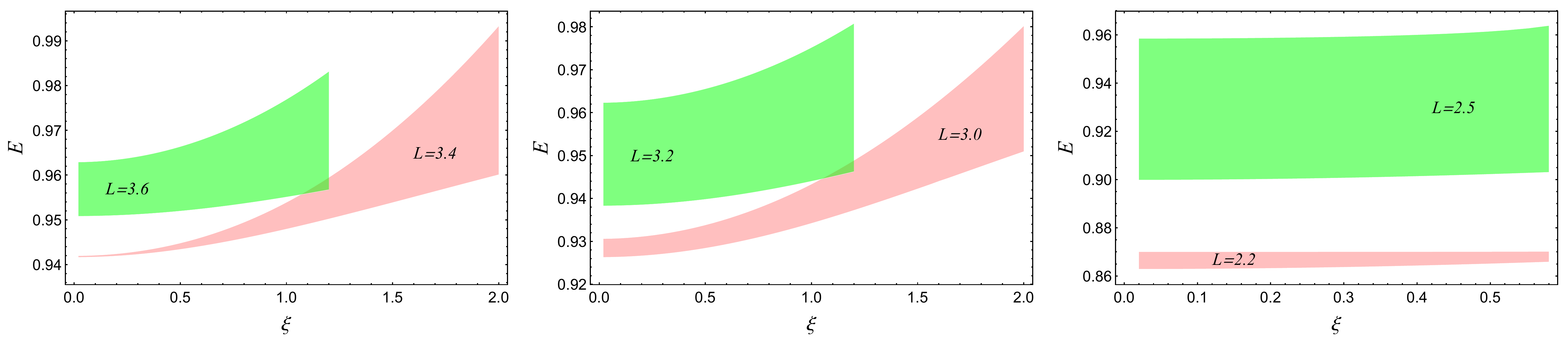}
    \includegraphics[width=\textwidth]{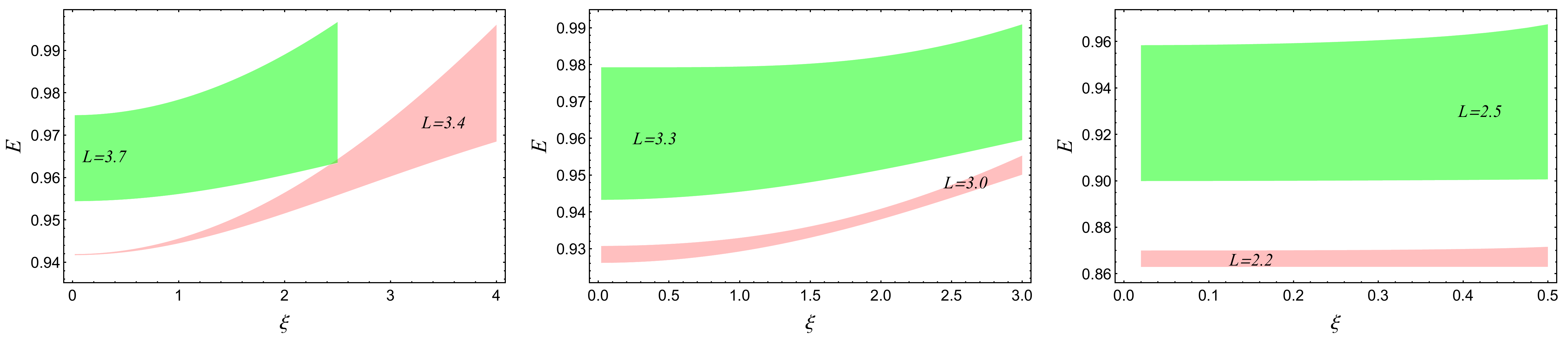}
    \caption{The allowed region of $E$ as the function of $\xi$ with selected $L$ for the prograde cases in the BH-I (Top) and BH-II (Bottom). From left to right, $a=0.1,\,0.5,\,0.9$.}
    \label{fig:BHPgammavsE}
\end{figure*}

Since the periastron $r_p$ and apastron $r_a$ are the two different roots of $\dot{r}=0$, the allowed value of either $E$ or $L$ will lie between the following  parameter region, i.e., $E_{ISCO}<E<E_{MBO}$ and $|L_{ISCO}|<|L|<|L_{MBO}|$. However, as illustrated in Fig. \ref{fig:MBO}-\ref{fig:ISCO-II},  we can not fix an unique $L$ for all $\xi$ when $a$ is large enough, hence we lack a valid way to check the influences of $\xi$ on the allowed range of $E$. In Fig. \ref{fig:BHPgammavsE} we show the allowed region of $E$ as the function of $\xi$ with selected $L$ for the prograde geodesic in the type BH-I and type BH-II. 

From these figures, it is evident that for systems with relatively small spin parameter $a$, when $\xi$ is sufficiently small, the closer the selected angular momentum $L$ approaches $L_{ISCO}$, the more restricted the allowable energy range for $E$ becomes; as $\xi$ increases, this permissible energy range correspondingly expands. However, for large spin parameter, the allowable range of $E$ remains relatively insensitive to increasing $\xi$, which is due to the inherently limited range of viable $\xi$ values in large $a$ cases. 
Such behaviors align with the established relationship between $L$ and $E$ as functions of $\xi$ for test particles in the ISCO and MBO orbits: for a fixed $L$ value, increasing $\xi$ drives the orbital parameters toward $L_{MBO}$. One can easily understand such behaviors from Eq. (\ref{eq:radialeq}) where $\dot{r}^2$ spans from $0$ to infinity. Geometrically, the requirement that $r_a$ and $r_p$ exist ensures that the curve of $\dot{r}^2$ must have at least one local maximum (corresponding to the MBO orbit) and two local minima, with the larger minimum corresponding to the ISCO orbit. These analysis reveal a clear correlation: proximity to the MBO orbit yields increasingly larger separations between $r_a$ and $r_p$, resulting in expanded energy selection ranges; conversely, proximity to the ISCO orbit produces diminishing separations between these values, thereby restricting the permissible energy range. 

The most significant difference between type BH-I and type BH-II is for the case $a=0.5$. For type BH-I, the allowed range $E$ resembles the aforementioned behavior, but  for type BH-II it is not the case. In fact, both $L_{ISCO}$ and $L_{MBO}$ of type BH-II decrease with increasing $\xi$ at a substantially more gradual rate compared to type BH-I. Consequently, within the physically permissible range of $\xi$, the selected angular momentum $L$ exhibits a similar relationship with both $L_{ISCO}$ and $L_{MBO}$ (see the $L_{MBO}$ of type BH-II in left bottom corner in Fig. \ref{fig:MBO} and $L_{ISCO}$ of type BH-II in right top corner in  Fig. \ref{fig:ISCO-II}). This geometric configuration results in only marginal variations in the allowable energy range for particle orbits. Furthermore, one can easily check that the similar behaviors exist for the retrograde orbits. 

\begin{figure*}
    \centering
    \includegraphics[width=\textwidth]{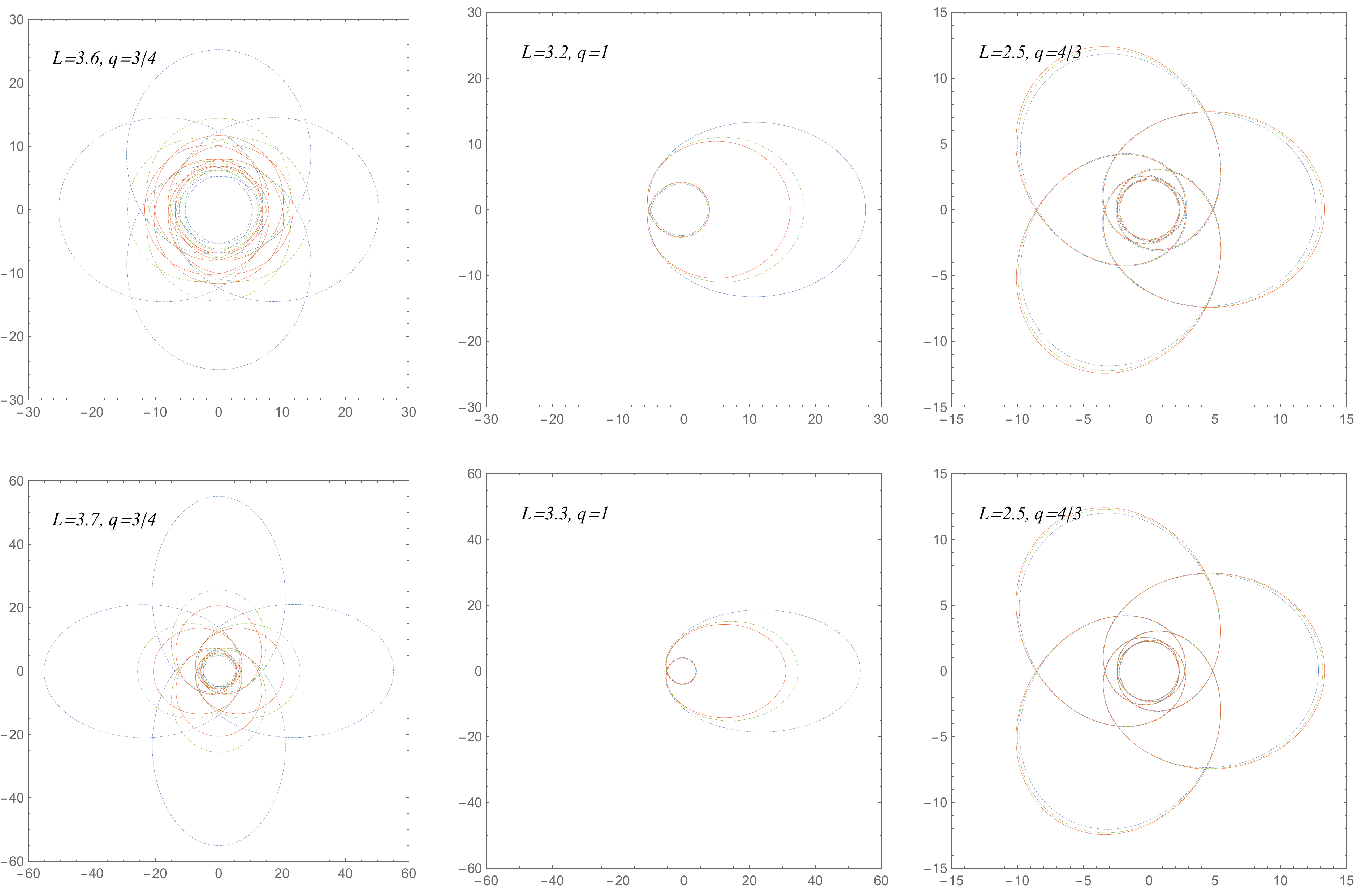}
    \caption{Periodic orbits with topology number $q$ (defined in Eq. 33) for BH-I (top row) and BH-II (bottom row). From left to right: $a = 0.1, 0.5, 0.9$. For each subfigure, three values of the regularization parameter $\xi$ are shown, decreasing from blue dotted to green dot-dashed to red solid lines (see legend). Note: The angular momentum $L$ and $\xi$ values differ between BH-I and BH-II for the same column because the admissible parameter ranges differ between the two black hole types; in each case, parameters are chosen to produce orbits of the same topology $q$ while remaining within the physically allowed region. For $a = 0.1$: BH-I uses $(L, \xi) = (3.6; 1, 0.5, 0.1)$ and BH-II uses $(L, \xi) = (3.7; 2, 1, 0.1)$. For $a = 0.5$: BH-I uses $(L, \xi) = (3.2; 1, 0.5, 0.1)$ and BH-II uses $(L, \xi) = (3.3; 2.5, 1.5, 0.5)$. For $a = 0.9$: both types use $(L, \xi) = (2.5; 0.5, 0.3, 0.1)$.}
    \label{fig:BHGW}
\end{figure*}

\begin{figure*}
    \centering
    \includegraphics[width=\textwidth]{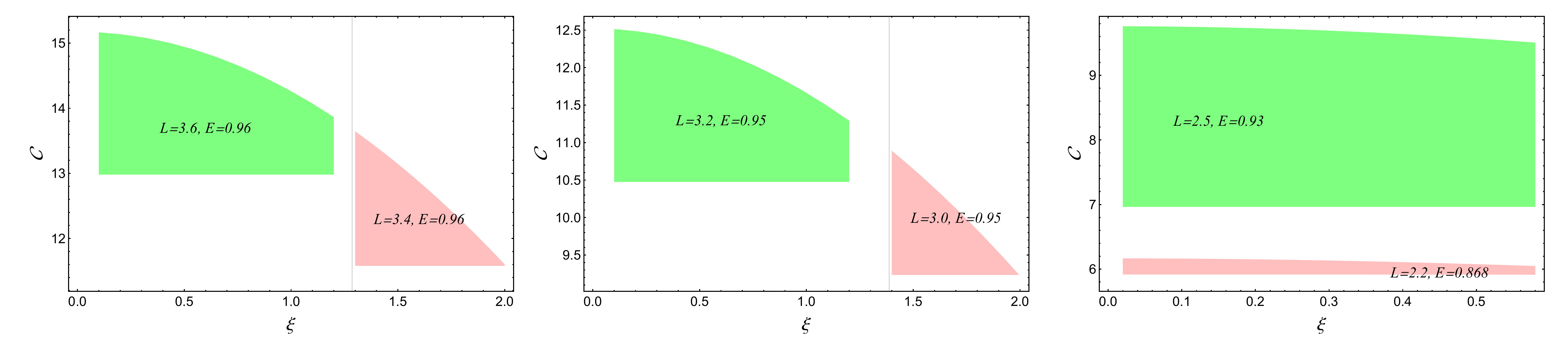}
    \caption{The allowed region of $\mathcal{C}$ as the function of $\xi$ with selected $L$ and $E$ for the prograde cases in the type BH-I. From left to right, $a=0.1,\,0.5,\,0.9$.}
    \label{fig:BHPgammavsc}
\end{figure*}

At the end of this section, we present a visual illustration of the influence of  the parameter $\xi$ on the periodic orbits in Fig. \ref{fig:BHGW}. Evidently, from the figures, the sensitivity of periodic orbits to $\xi$ diminishes as the spin parameter $a$ increases, which is fully consistent with the  conclusions presented above (see Figs. \ref{fig:MBO}, \ref{fig:ISCO-I}, \ref{fig:ISCO-II}). For different types of black holes, the orbits are almost the same when the spin parameter $a$ approaches unity. Meanwhile, for the same $a$, we will obtain large orbits as $\xi$ grows large, because the selected angular momentum $L$ is close to $L_{MBO}$ for large $\xi$, leading to a large $E$ and finally resulting in a large orbit. Notably, when the spin parameter approaches unity, larger values of $\xi$ lead to a reduction in the apastron radius of the periodic orbit, in sharp contrast to the behaviors observed in the low-spin regime.

\subsection{General cases}

We now provide a brief discussion on the effects of the parameter $\xi$ on the general bound orbits of the test particles outside the equatorial plane, leaving a more detailed exploration for future work. For the sake of simplicity, we restrict our discussion of the general test geodesics to the meridian plane generated by $r$ and $\theta$. 

According to Eqs.(\ref{eq:geodesiceq5},\,\ref{eq:geodesiceq6}), the Carter constant $\mathcal{C}$ reaches its minimum value $\mathcal{C}_{0}=a^2E^2+L^2$ on the equatorial plane. For both type BH-I and BH-II, one has
\begin{eqnarray}
    \lim_{r\rightarrow0}\mathcal{R}(r)&=&\frac{4M^2\xi^2(2aL-\mathcal{C})}{r^2}\rightarrow-\infty,\\
    \lim_{r\rightarrow\infty}\mathcal{R}(r)&=&r^4(E^2-1)\rightarrow-\infty.
\end{eqnarray}
The first relation holds because the value of $L$ is already constrained by the condition $\lvert{L}\rvert\in(\lvert{L_{ISCO}}\rvert,\,\lvert{L_{MBO}}\rvert)$. So, to ensure the existence of the general bound orbits, $\mathcal{R}(r)$ must have at least one local minimum. This inevitably imposes new constraints on the Carter constant. Therefore, by setting $\mathcal{R}(r)=0$ and $\mathcal{R}'(r)=0$, one can solve for the critical value $\mathcal{C}_{c}$ and its corresponding radius $r_{c}^{\mathcal{C}}$, both of which are functions of the parameters $(L,\,E\,,\xi,\,a)$. 

As previously established, once $\xi\,,a$ and $L$ are specified, the allowable range of $E$ can be determined. Choosing an appropriate value of $E$ within this range then allows the computation of $\mathcal{C}_{c}$ and $r_{c}^{\mathcal{C}}$. Moreover, since $\mathcal{R}(r)$ possesses at least one local minimum, $\mathcal{C}$ will be bounded from above by $\mathcal{C}_{max}$ and from below by $\mathcal{C}_{min}$, i.e., $\mathcal{C}_{c}=\mathcal{C}_{max(min)}$. Comparing the lower bound with $\mathcal{C}_{0}$ ultimately yields the admissible range for $\mathcal{C}\in[\mathcal{C}_{max},\, \,\text{min}\left(\mathcal{C}_{min}\,,\mathcal{C}_{0}\right)]$. In Fig. \ref{fig:BHPgammavsc} we plot the allowable range of the Carter constant $\mathcal{C}$ with fixed $L,\, E$ as a function of $\xi$ for type BH-I with the prograde orbits, showing that the upper bound $\mathcal{C}_{max}$ decreases with $\xi$, such behavior also apply for the retrograde orbits in both types of black holes.

\begin{figure}
    \centering
    \includegraphics[width=0.45\textwidth]{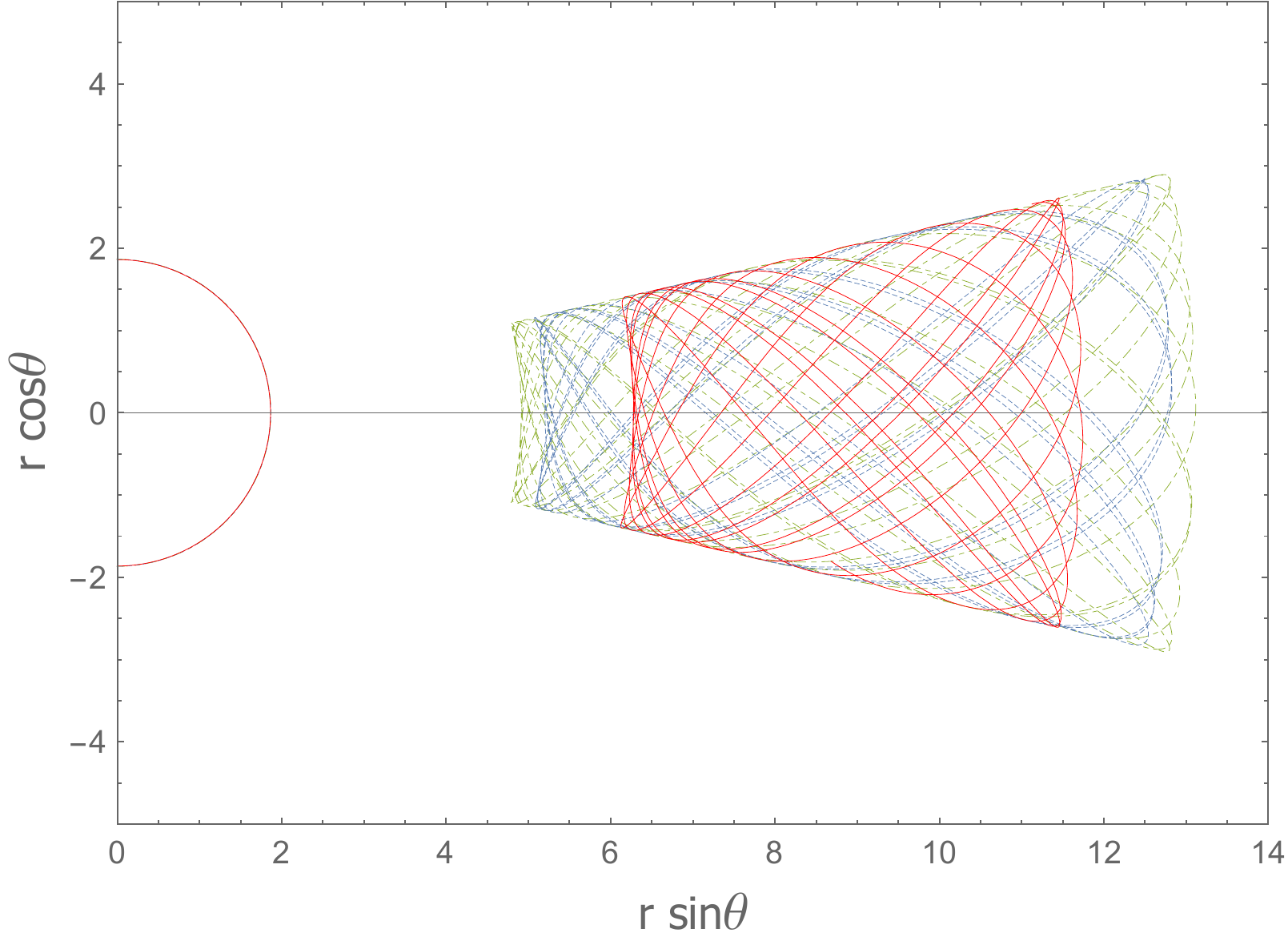}
    \caption{The projections of prograde orbits on the meridian plan for $a=0.5\,,L=3.2,\,E=0.95,\,\mathcal{C}=11$ of type BH-I. The red solid line, blue dotted line and green dashed line correspond to $\xi=1,\,0.5\,,0.1$, respectively. The half circles describe the horizon. Note for different $\xi$ the horizon radius are almost same.}
    \label{fig:BHImeridian}
\end{figure}

Finally, as an example,  we plot the projections of orbits on the meridian plane $(r\sin{\theta}, \,r\cos{\theta})$ with selected parameters $(L,\,E,\,\mathcal{C},\,\xi,\,a=0.5)$ in Fig. \ref{fig:BHImeridian}  for type BH-I . Clearly, with large $\xi$, the range of the particle orbit radius becomes narrower, and such behaviors also hold for other cases.

\section{The Gravitational wave from periodic orbits}\label{sec:GW}

In this section, we will provide a preliminary exploration of the gravitational waveforms emitted by the periodic orbits of a test particle orbiting a supermassive LQG black hole, by assuming that the test particle has a mass extremely smaller than the central black hole and moves on the equatorial plan. Moreover, focusing on the periodic orbits, we  neglect the energy loss due to gravitational radiation from the secondary body and employ the leading order post-Newtonian approximation to compute the gravitational waveform, thereby isolating and examining the influence of the parameter $\xi$. Although this simplified model does not capture the full relativistic dynamics of EMRI systems, it yields useful qualitative insights into how the parameter $\xi$ imprints on the waveform morphology. A more realistic treatment which considers the radiative energy loss and employs techniques such as numerical kludge method is beyond the scope of the present study and will be addressed in future work.

Based on the given toy model and following Refs. \cite{Maselli:2021men,Liang:2022gdk,Babak:2006uv,Poisson:2014kt,Chua:2017ujo}, the gravitational wave radiated from the periodic orbits can be computed using the quadratic order formula:
\begin{eqnarray}\label{GWfirst}
    h_{ij}=\frac{2G}{c^4D_{L}}\ddot{I}_{ij}.
\end{eqnarray}
{Here, $I_{ij}=\int d^{3}xT^{tt}(t,x^i)x^ix^j$ is the mass quadrupole moment given in terms of the source stress-energy tensor $T^{tt}(t,x^i)=m\delta^{3}(x^i-z^{i}(t))$, with the Cartesian spatial coordinates $x^i,\,z^i(t)$  defined in Ref. \cite{Babak:2006uv}. For a binary EMRI system of orbiting bodies, Eq. (\ref{GWfirst}) becomes 
\begin{eqnarray}
    h_{ij}=\frac{4G\mu}{c^{4}D_{L}}\left(v_{i}v_{j}-\frac{G(M+m)}{r}n_{i}n_{j}\right),
\end{eqnarray}
where $\mu=Mm/(M+m)$ is the symmetric reduce mass with $M,\,m$ the masses of the LQG black hole and the secondary object, respectively. $D_{L}$ is the luminosity distance of the EMRI system, $\boldsymbol{v}$ is the spatial relative velocity of the secondary object and $\boldsymbol{n}$ is the unit vector pointing to the radial direction associated to the motion of the secondary object \cite{Poisson:2014kt}. }

\begin{figure*}
    \centering
    \includegraphics[width=\textwidth]{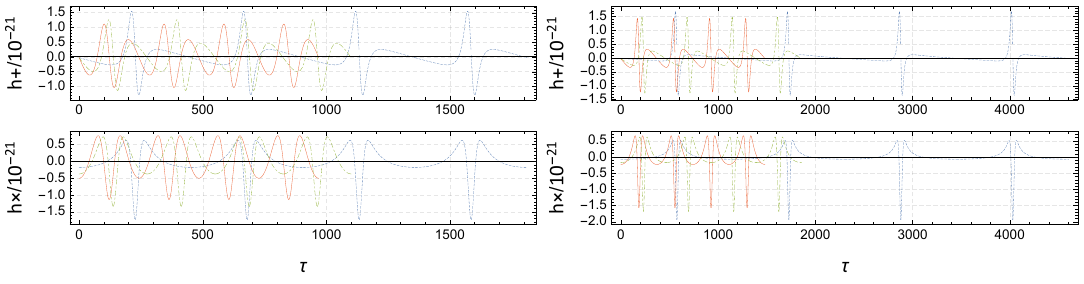}
    \caption{The gravitational waveforms of the prograde periodic bound orbits correspond to type BH-I with $L=3.6,\,q=3/4,\,a=0.1$ (left column) and type BH-II with $L=3.7,\,q=3/4,\,a=0.1$ (right column). The total accumulated azimuth is $\Delta\phi=14\pi$ for each line.}
    \label{fig:BHGW2}
\end{figure*}

Now, to describe the orbital motion, we can define an orbital eccentricity $e$ and semi-latus rectum $p$ using the conventional Keplerian definitions
\begin{eqnarray}\label{eq:rpra}
    r_p=\frac{p}{1+e},\qquad r_{a}=\frac{p}{1-e},
\end{eqnarray}
then the orbits can be described in terms of a new angular variable $\chi$ as
\begin{eqnarray}
    r=\frac{p}{1+e\cos{\chi}}.
\end{eqnarray}
Clearly, as the parameter $\chi$ varies from 0 to $2\pi$, the coordinate $r$ goes back and forth between $r_{a}$ and $r_p$. We can then introduce a ``orbit-adapted'' system $(x,y,z)$ in the orbital plane as follows: setting the origin as the barycenter, then $(x-y)$ plane coincides with the orbital plane and the $z$-axis points in the direction of the angular momentum vector, $x$-axis points to the apastron. In this frame, the associated unit vectors are given by
\begin{eqnarray}
  \bm{n}=\left(\cos{\chi},\,\sin{\chi},\,0\right),\quad \bm{\lambda}=\left(-\sin{\chi}\,,\cos{\chi},\,0\right)
\end{eqnarray}
and then
\begin{eqnarray}
    \bm{r}=r\bm{n},\qquad \bm{v}=\dot{r}\bm{n}+r\dot{\chi} \bm{\lambda}.
\end{eqnarray}
Based on the orbit adapted coordinates, to better describe the gravitational waveform from the perspective of a distant observer, we can further introduce a ``detector-adapted'' Cartesian coordinate system $(X, Y, Z)$ as
\begin{eqnarray}
    \boldsymbol{e}_{X}&=&(\cos{\zeta},\,-\sin{\zeta},\,0),\\
    \boldsymbol{e}_{Y}&=&(\cos{\iota}\sin{\zeta},\,\cos{\iota}\cos{\zeta},\,-\sin{\iota}),\\
    \boldsymbol{e}_{Z}&=&(\sin{\iota}\sin{\zeta},\,\sin{\iota}\cos{\zeta},\,\cos{\iota}),
\end{eqnarray}
{where $\iota,\,\zeta$ denotes the inclination angle between the EMRI’s orbital angular momentum and the line of sight and the latitudinal angle between the pericenter and the line of nodes, as measured in the orbital plane, respectively. Such system has the same origin as the orbit system $(x, y, z)$, and the $Z$-axis points in the direction of the gravitational wave detector. The $(X-Y)$ plane is orthogonal to the $Z$-axis and coincides with the plane of the sky from the detector’s point of view, and the $X$-axis points toward the ascending node, the point at which the orbit cuts the plane from below \cite{Poisson:2014kt}. Then, by adopting $\bm{e}_{X}$ and $\bm{e}_{Y}$ as the vectorial basis in the transverse subspace, the gravitational wave polarizations, $h_{+}$ and $h_{\times}$, take the forms}
\begin{eqnarray}\label{GWpolar}
    h_{+}&=&\frac{1}{2}\left(e_{X}^{j}e_{X}^{k}-e_{Y}^{j}e_{Y}^{k}\right)h_{jk},\\
    h_{\times}&=&\frac{1}{2}\left(e_{X}^{j}e_{Y}^{k}+e_{Y}^{j}e_{X}^{k}\right)h_{jk}.
\end{eqnarray}
{Inserting the expressions of $\bm{n}$ and $\bm{\lambda}$ in the detector-adapted coordinates
\begin{eqnarray}
\bm{n}&=&\left[\cos{ \tilde{\zeta}},\,\cos{\iota}\sin{ \tilde{\zeta}}\,,\sin{\iota}\sin{ \tilde{\zeta}}\right],\\
    \bm{\lambda}&=&\left[-\sin{ \tilde{\zeta}},\,\cos{\iota}\cos{ \tilde{\zeta}}\,,\sin{\iota}\cos{ \tilde{\zeta}}\right],\\
    \tilde{\zeta}&=&\zeta+\chi,
\end{eqnarray}
within Eq. (\ref{GWpolar}), we have
\begin{eqnarray}
    h_{+}&=&-h_0(1+\cos^{2}{\iota})\left[\cos{(2\chi+2\zeta)+\frac{5}{4}e\cos{}(\chi+2\zeta)}\right.\nonumber\\
    &\quad&+\left.\frac{1}{4}e\cos{(3\chi+2\zeta)}+\frac{1}{2}e^2\cos{2\zeta}\right] \nonumber\\
    &\quad&+\frac{1}{2}e\sin^{2}{\iota}(\cos{\chi}+e),  \\
    h_{\times}&=&-2h_0\cos{\iota}\left[\sin{(2\chi+2\zeta)}+\frac{5}{4}e\sin{(\chi+2\zeta)}\right.\nonumber \\
    &\quad&+\left.\frac{1}{4}e\sin{(3\chi+2\zeta)}+\frac{1}{2}e^2\sin{2\zeta}\right],\\
    h_0&=&\frac{2G_0}{c^2D_{L}p}\frac{m}{M}.
\end{eqnarray}}

In Fig. \ref{fig:BHGW2}, we present the influence of $\xi$ on the gravitational waveforms of the periodic orbits of the secondary object with the parameters given by the first column in Fig. \ref{fig:BHGW}. Here, we have considered an EMRI system that consists of a supermassive LQG black hole with mass $M=10^{7} M_{\odot}$ and a secondary object with mass $m=10M_{\odot}$. The other parameters have been set as $D_{L}=200$ Mpc, $\iota=\zeta=\pi/4$. Clearly,the figures convey information that is consistent  with that presented in Fig. \ref{fig:BHGW}. For both black hole types, a larger value of $\xi$ leads to a more pronounced effect on the gravitational waveform. Notably, for type BH-II, the influence of $\xi$ is significantly more prominent compared to that for type BH-I. 

Our results show qualitatively similar $\xi$-dependent enhancements of gravitational wave amplitude as in Refs. \cite{Ahmed:2025shr,Yang:2024lmj} in the static case, but slightly different from results  in Ref. \cite{Chen:2025aqh} at first sight, however, after rescaling the metric (\ref{eq:rmetric})  by multiplying the conformal factor $\sqrt{f(r)/g(r)}$  and taking the limit $a\rightarrow0$, we obtain same results as \cite{Chen:2025aqh}.

\begin{figure*}
    \centering
    \includegraphics[width=\textwidth]{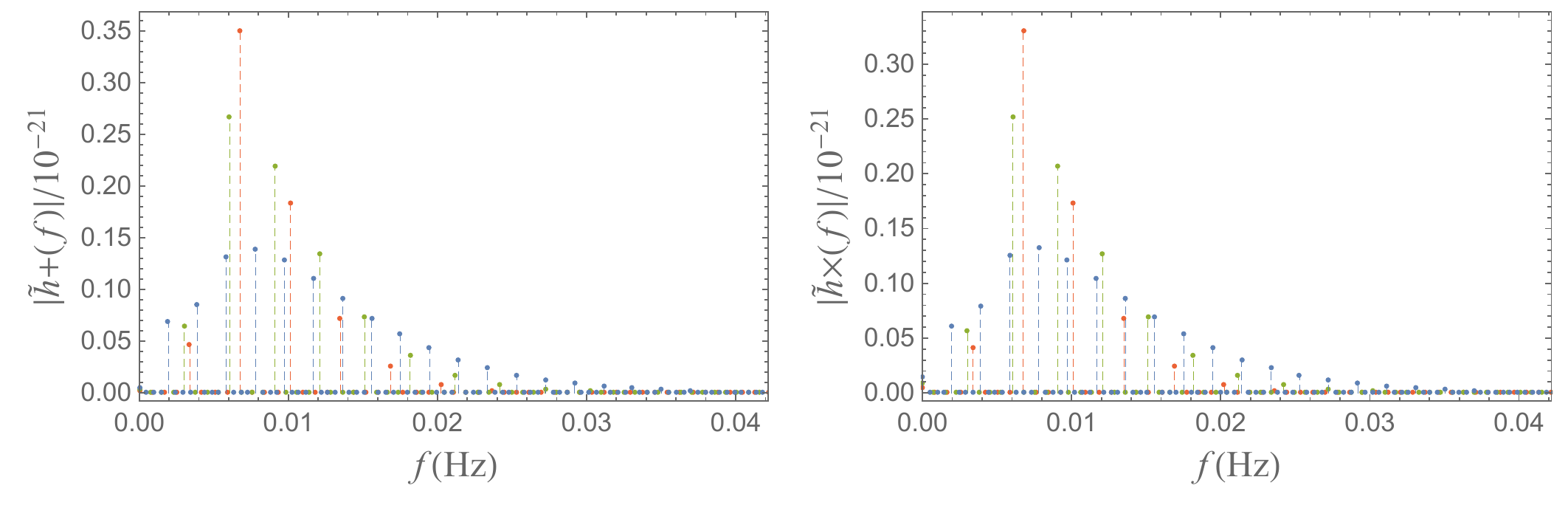}
    \caption{The absolute frequency spectra $\lvert \tilde{h}_{+}(f)\rvert$ and $\lvert \tilde{h}_{\times}(f)\rvert$ corresponding to the gravitational waveforms of BH-I from Fig. \ref{fig:BHGW2}. The blue dotted line, the green dotdashed line and the red solid line correspond to $\xi=(1,\,0.5\,,0.1)$, respectively.}
    \label{fig:freBHI}
\end{figure*}

\begin{figure*}
    \centering
    \includegraphics[width=\textwidth]{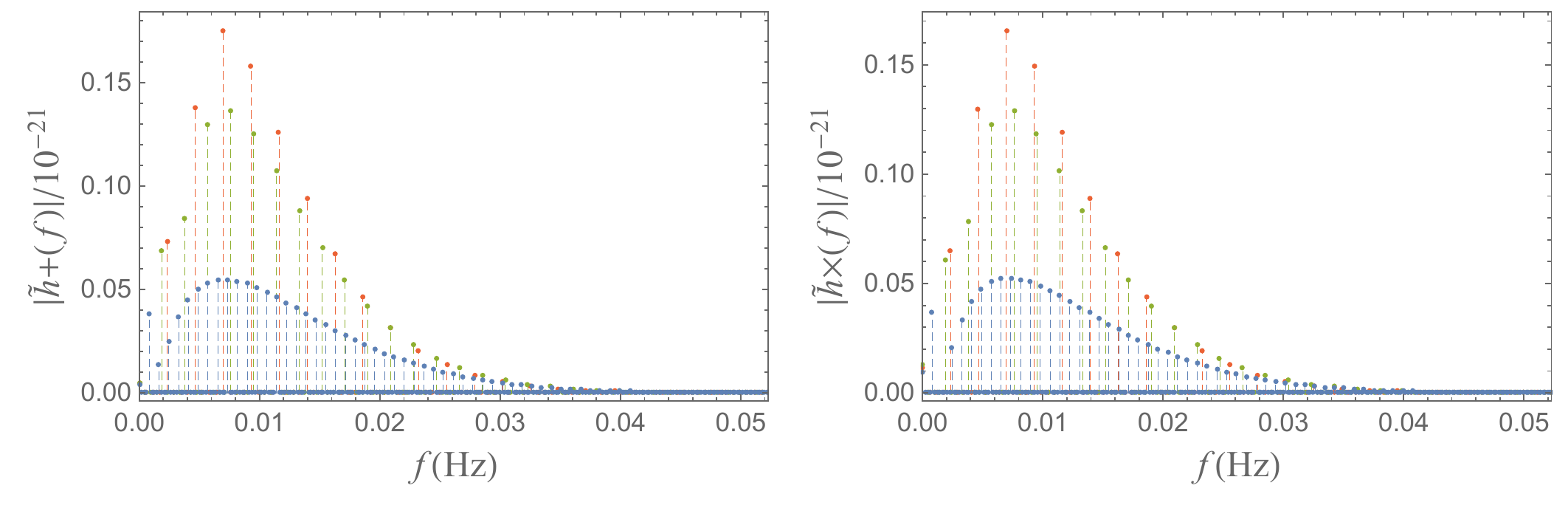}
    \caption{The absolute frequency spectra $\lvert \tilde{h}_{+}(f)\rvert$ and $\lvert \tilde{h}_{\times}(f)\rvert$ corresponding to the gravitational waveforms of BH-II from Fig. \ref{fig:BHGW2}. The blue dotted line, the green dotdashed line and the red solid line correspond to $\xi=(2,\,1\,,0.1)$, respectively.}
    \label{fig:freBHII}
\end{figure*}

\begin{figure*}
    \centering
    \includegraphics[width=0.49\textwidth]{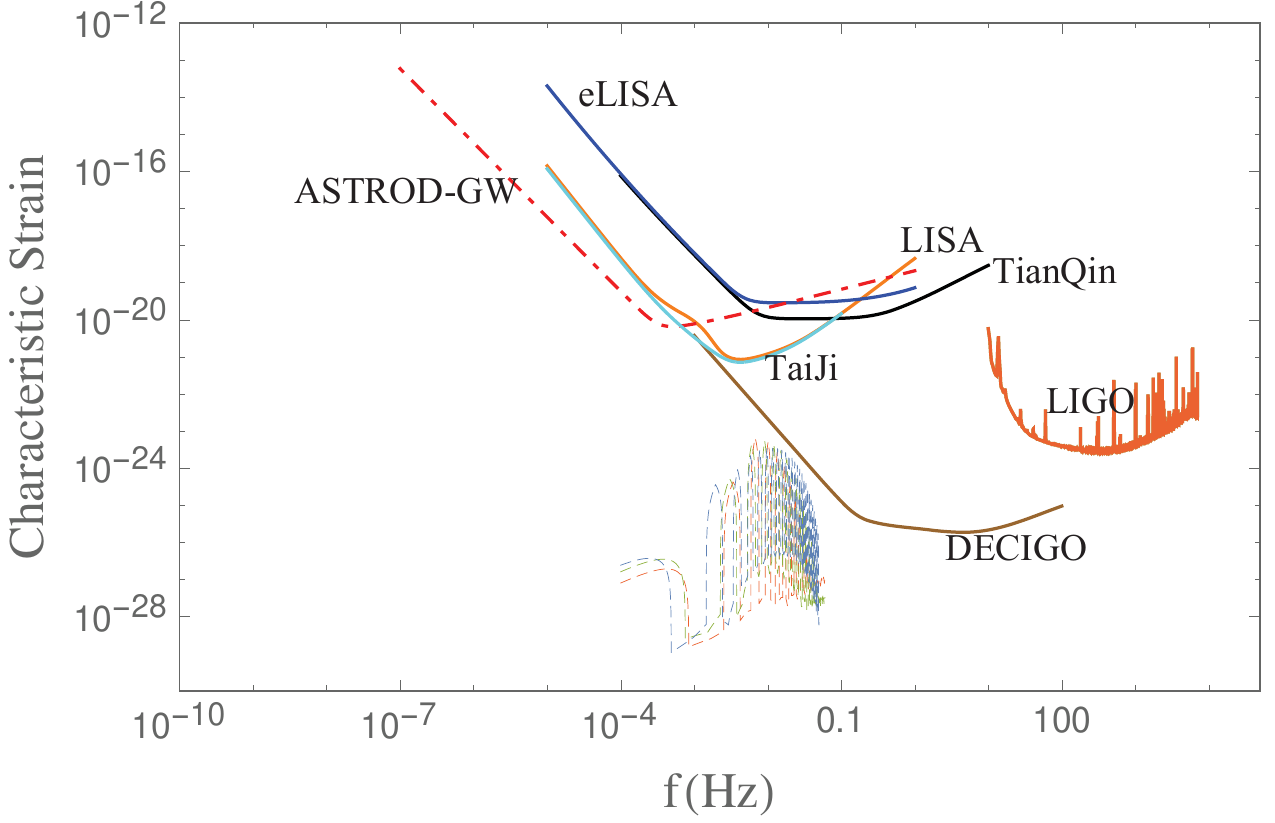}
    \includegraphics[width=0.49\textwidth]{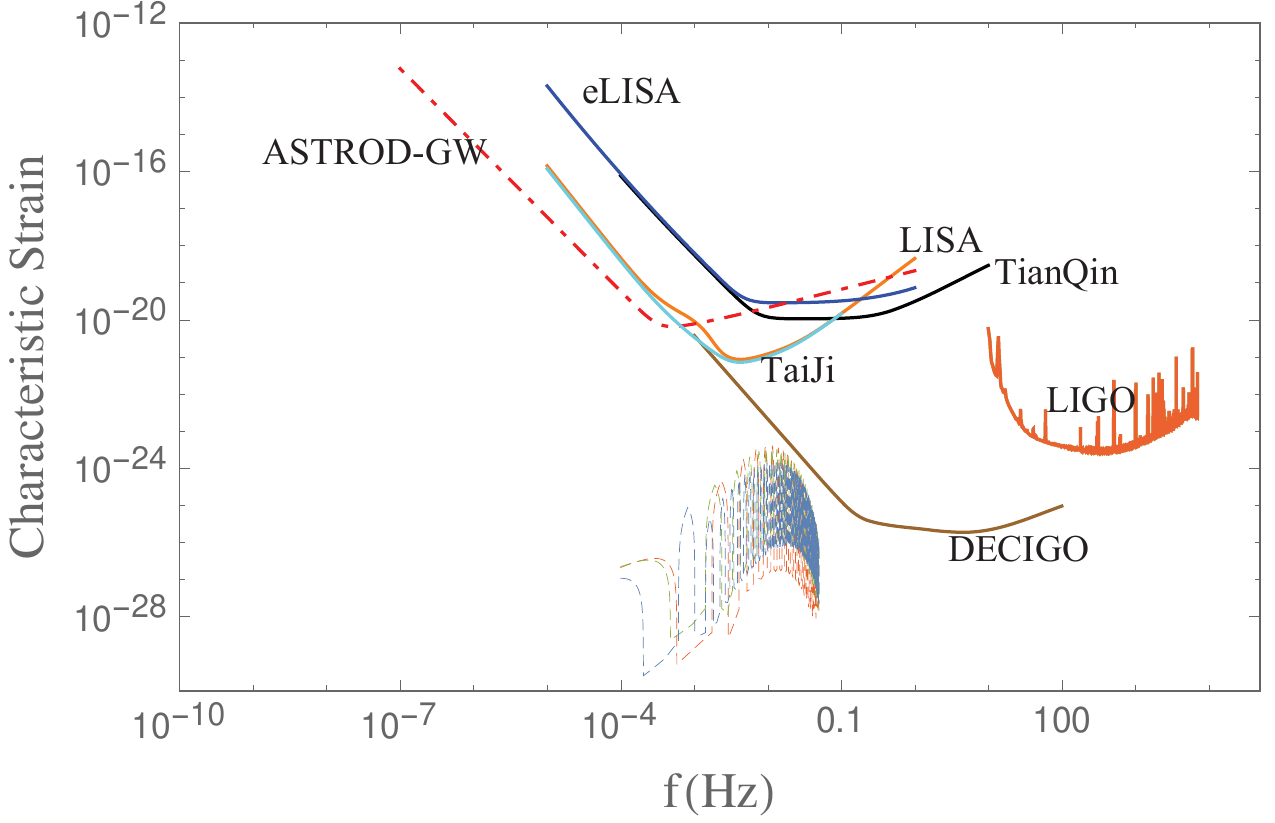}
    \caption{Comparison of the characteristic strain of gravitational waves emitted from the periodic orbits around BH-I (left) and BH-II (right)  to the sensitivity curves of various detectors. The parameters are inherited from Fig. \ref{fig:BHGW2}.}
    \label{fig:StrainBH}
\end{figure*}

The gravitational waveforms generated by the secondary object on the periodic orbital motions can be further analyzed through the corresponding frequency spectra $\lvert \tilde{h}_{+}(f)\rvert$ and $\lvert \tilde{h}_{\times}(f)\rvert$ by applying a discrete Fourier transform (DFT) to the time-domain gravitational waveforms $h_+$ and $h_{\times}$. Such spectra provide a detailed examination of the signal's frequency distribution and are adopted to define the sensitivity curve of the detectors \cite{Creighton:2011zz,Barack:2018yly,Ni:2024acg,Yunes:2025xwp}. This curve specifies the detector’s equivalent noise level across specific Fourier frequency ranges, with the vertical axis typically represented in logarithmic coordinates of amplitude spectral density or so-called the characteristic strain, defined by
\begin{eqnarray}
    S_{c}(f)=2f\sqrt{(\lvert \tilde{h}_{+}(f)\rvert^{2}+\lvert \tilde{h}_{\times}(f)\rvert^2)},
\end{eqnarray}
Adopting the parameters from Fig. \ref{fig:BHGW2} we first show the absolute frequency spectra $\lvert \tilde{h}_{+}(f)\rvert$ and $\lvert \tilde{h}_{\times}(f)\rvert$ of the gravitational waves in Figs. \ref{fig:freBHI}, \ref{fig:freBHII} and then compare the characteristic strain $S_{c}(f)$ of the type BH-I and type BH-II with the sensitivity curves of various space or ground based detectors, such as LISA $\&$ eLISA \cite{Amaro-Seoane:2012vvq,LIGOScientific:2014pky,LISA:2022kgy,LIGOScientific:2007fwp}, TianQin \cite{TianQin:2015yph,Torres-Orjuela:2023hfd,Li:2024rnk}, TaiJi \cite{Hu:2017mde,Luo:2019zal,Gong:2021gvw,Liu:2023qap}, LIGO \cite{Essick:2025zed,LIGOScientific:2025slb}, ASTROD-GW \cite{Ni:2012eh} or DECIGO \cite{Ishikawa:2020hlo}, in Fig. \ref{fig:StrainBH}. This visual comparison shows that the characteristic frequencies of the periodic orbits $(q=3/4,\,a=0.1)$ on equatorial planes for both black hole types are concentrated between $(10^{-3},0.1)$ Hz but the corresponding characteristic strains are beyond the range of most detectors.

\section{The evolutions of the semi-latus rectum for near-circular orbits}\label{sec:evop}

\begin{figure*}
    \centering
    \includegraphics[width=\textwidth]{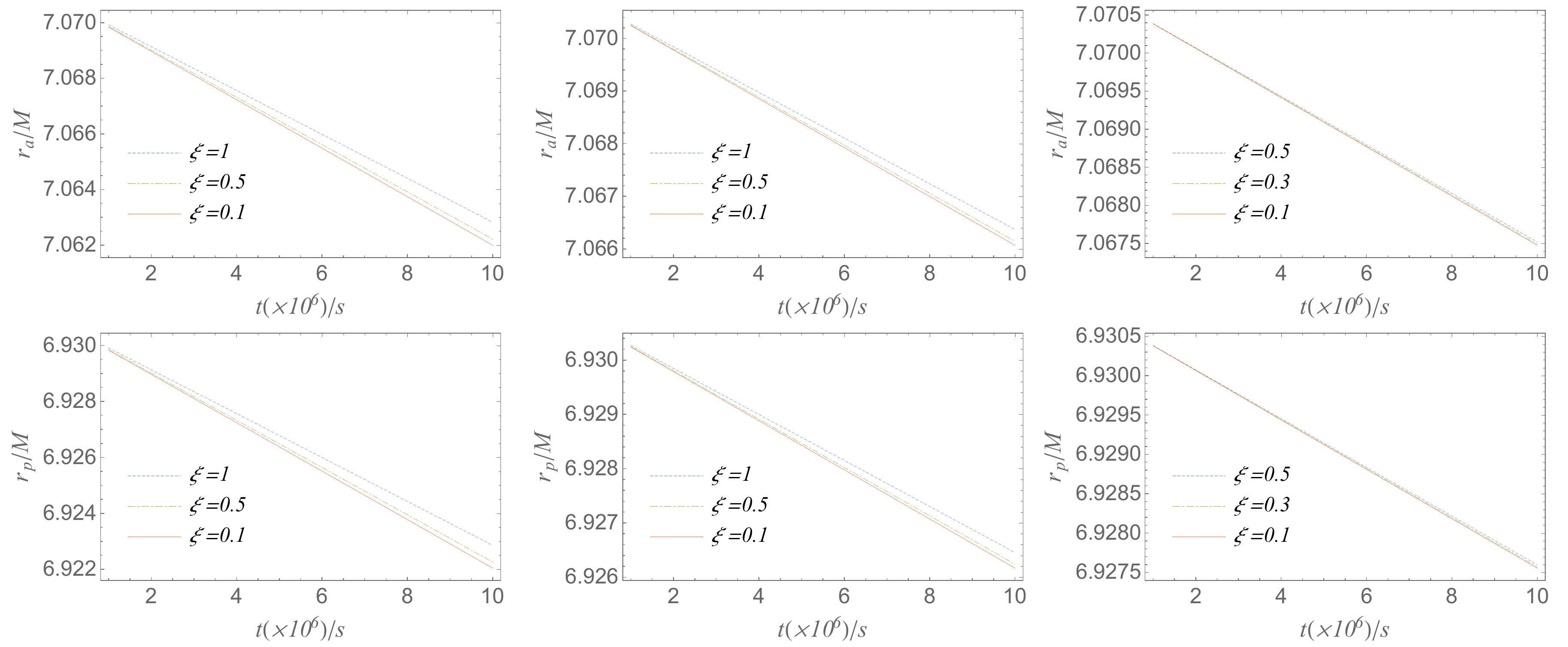}
    \caption{The evolution of apastron $r_{a}/M$ and $r_{p}/M$ for the near-circular orbits of BH-I, where the initial value $p(t_0=0)=7,\,e(t_0=0)=10^{-2}$. From left to right, $a=0.1,\,0.5,\,0.9$.}
    \label{fig:pBHI}
\end{figure*}

\begin{figure*}
    \centering
    \includegraphics[width=\textwidth]{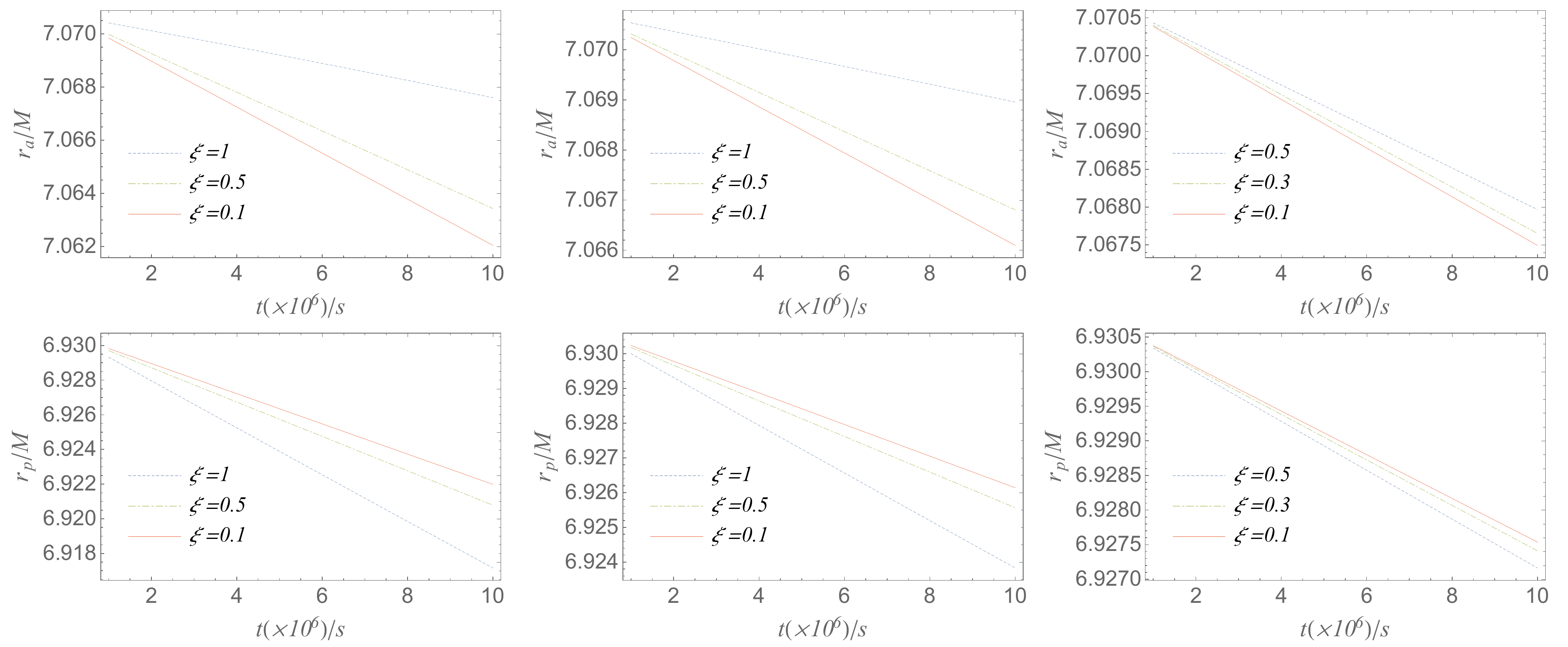}
    \caption{The evolution of apastron $r_{a}/M$ and $r_{p}/M$ for the near-circular orbits of BH-II, where the initial value $p(t_0=0)=7,\,e(t_0=0)=10^{-2}$. From left to right, $a=0.1,\,0.5,\,0.9$.}
    \label{fig:pBHII}
\end{figure*}

In the previous section we have focused on the periodic orbits of the secondary object without considering the energy loss from the gravitational radiation. In this section we will derive orbital evolution in the LQGBH background, which is primarily driven by the energy and angular momentum flux caused by the gravitational radiation. However, due to the extreme complexity of calculations, we consider only the near-circular orbits on the equatorial plane, i.e., $e\ll1$ and we concern only the first order of $e$. For the energy and angular momentum flux we adopt the expressions given in Refs. \cite{Peters:1963ux,Peters:1964zz,Ryan:1995zm,Maggiore:2007ulw,Liu:2024qci,Zhao:2025sck,Gong:2025mne,Muguruza:2026hqn} using the standard quadrupole approximation as
\begin{eqnarray}
    \left\langle\frac{dE}{dt}\right\rangle&=&-\frac{1}{5}\left\langle\frac{d^{3}Q_{ij}}{dt^3}\frac{d^3Q^{ij}}{dt^3}-\frac{1}{3}\frac{d^3Q_{ii}}{dt^3}\frac{d^{3}Q^{jj}}{dt^3}\right\rangle,\label{eq:aveEn}\\
    \left\langle\frac{dL_{i}}{dt}\right\rangle&=&-\frac{2}{5}\epsilon_{ijk}\left\langle\frac{d^{2}Q_{jm}}{dt^2}\frac{d^{3}Q^{km}}{dt^3}\right\rangle,\label{eq:aveL}
\end{eqnarray}
where $Q^{ij}=\mu x^{i}x^{j}$ is the inertia tensor and $x^{i}=(r\cos{\phi},\,r\sin{\phi},\,0)$ is the position vector between the black hole and the secondary object represented in spherical coordinates on the equatorial plane. The average is taken over one radial period. Under the adiabatic approximation the changes in the orbital energy and angular momentum of the secondary object are then completely converted into the energy radiated as gravitational waves, i.e.,
\begin{eqnarray}
    \frac{dE_{GW}}{dt}&=&\left\langle\frac{dE}{dt}\right\rangle=\frac{dE_{ob}}{dt},\\
    \frac{dL_{GW}}{dt}&=&\left\langle\frac{dL_{z}}{dt}\right\rangle=\frac{dL_{ob}}{dt}.
\end{eqnarray}
Hereafter, we will drop the subscript “ob” without causing any confusion. For general bounded orbits on the equatorial plane, the expressions of $E$ and $L$ should be the functions of $p$ and $e$ according to definitions (\ref{eq:radialeq}, \ref{eq:rpra}). However, since we only consider the near-circular orbits so $E=E_{1}(p)+E_{2}(p)e^{2}+\mathcal{O}(e^3)$ and $L=L_{1}(p)+L_{2}(p)e^{2}+\mathcal{O}(e^3)$, where the expressions for $E_{1(2)}(p)$ and $L_{1(2)}(p)$ are extremely complicated. Substituting (\ref{eq:geodesiceq1}), (\ref{eq:geodesiceq2}), (\ref{eq:geodesiceq3}), (\ref{eq:geodesiceq4}) into (\ref{eq:aveEn}) and (\ref{eq:aveL}), one will get
\begin{eqnarray}
    \frac{dE}{dt}&=&-\frac{32\mu^2p^{4}}{5}\phi^{6}_{1}+\mathcal{O}(e^{2}),\label{eq:evoE1}\\
    \frac{dL}{dt}&=&-\frac{32\mu^{2}p^4}{5}\phi^{5}_{1}+\mathcal{O}(e^2),\label{eq:evoL1}
\end{eqnarray}
where $\phi_{1}=\phi_{1}(p)$ is the first term after expanding $\phi'(t)$ around $e=0$, i.e., $\phi'(t)=\phi_{1}(p)+\phi_{2}(p)e\cos{\chi}+\mathcal{O}(e^2)$. It is worth noting that the second term $\phi_{2}(p)e\cos{\chi}$ in the above expanded expression disappears after the averaging operation is performed. For BH-I,
\begin{eqnarray}
    \phi_{1}(p)=\frac{1}{a+p^{3}[Mp^3-(8M^2-6Mp+p^2)\xi^2]^{-1/2}},
\end{eqnarray}
while for BH-II
\begin{eqnarray}
    \phi_{1}(p)&=&\frac{aE_{1}(\mathcal{N}_{1}-\mathcal{N}_{2})+L_{1}\mathcal{N}_{2}}{E_{1}(\mathcal{N}_{3}+2a^2\mathcal{N}_{2})-aL_{1}(\mathcal{N}_{1}-\mathcal{N}_{2})}, \\
    \mathcal{N}_{1}&=&\sqrt{p^8+p^5(p-2M)\xi^2},\\
    \mathcal{N}_{2}&=&(p-2M) [p^3+(p-2M)\xi^2],\\
    \mathcal{N}_{3}&=&[p^3-a^2(p-2M)][p^3+(p-2M)\xi^2].
\end{eqnarray}
One can easily check that for $a\rightarrow0,\,\xi\rightarrow0$, Eqs. (\ref{eq:evoE1}, \ref{eq:evoL1}) reduce to the classical Schwarzschild case in GR. Up to first order of $e$, the evolutions of the semi-latus rectum $p(t)$ and the eccentricity $e(t)$ is described by
\begin{eqnarray}
    \frac{dp}{dt}&=&\frac{L_{2}\dot{E}-E_{2}\dot{L}_{z}}{\mathcal{N}_{}4},\\
    \frac{de}{dt}&=&\frac{E_{1}'\dot{L}_{z}-\dot{E}L_{1}'}{2e\mathcal{N}_{4}}+\frac{\mathcal{N}_{5}\mathcal{N}_{6}}{2\mathcal{N}_{4}^{2}}e,\\
    \mathcal{N}_{4}&=&L_{2}E_{1}'-E_{2}L_{1}',\\
    \mathcal{N}_{5}&=&L_{2}\dot{E}-E_{2}\dot{L}_{z},\\
    \mathcal{N}_{6}&=&E_{2}'L_{1}'-E_{1}'L_{2}',
\end{eqnarray}
where we again have ignored the $\mathcal{O}(e^2)$ terms. Note, here the dot and the prime stand for the derivative with respect to the coordinate time $t$ and the semi-latus rectum $p$, respectively. In Figs. \ref{fig:pBHI} and \ref{fig:pBHII} we plot the late-stage evolution of $r_{a}$ with the initial value $p(0)=7,\,e(0)=10^{-2}$ for different $a$ and $\xi$ under the prograde near-circular orbits approximation.
 
As illustrated in the figures, for both black hole configurations, the decay rate of $r_a$ exhibits a monotonic decrease with increasing spin parameter $a$, indicating that rapid rotation suppresses the dissipation of the apastron distance. Consequently, the influence of $\xi$ on the orbital decay dynamics becomes progressively attenuated with larger $a$, which is fully consistent with the conclusions established in the preceding sections. Notably, for both black hole types, when the spin parameter $a$ is held fixed, a larger value of $\xi$ corresponds to a systematically reduced decay rate of $r_a$, suggesting that the quantum gravitational correction acts to stabilize the apastron against orbital shrinkage. An analogous trend is observed for the periastron distance $r_p$ in the case of BH-I, where increasing $\xi$ similarly retards the decay of $r_p$. However, for BH-II, the dependence of $r_p$ on $\xi$ is qualitatively reversed, i.e., a larger $\xi$ gives rise to an enhanced decay rate of $r_p$ while $r_a$ remains comparatively stable, leading to a progressive increase in orbital eccentricity. This implies that the near-circular orbits in the BH-II spacetime undergo a transition toward more eccentric configurations under stronger quantum corrections, a behavior that sharply distinguishes BH-II from its BH-I counterpart and may leave observable imprints on the gravitational waveform morphology. This behavior is also consistent with Fig. \ref{fig:BHGW}. Furthermore, the value of $\xi$ must be selected to conform to the constraints illustrated in Figs. \ref{fig:MBO}, \ref{fig:ISCO-I}, \ref{fig:ISCO-II} when determining the initial value $p(0)$.

\section{Conclusions}\label{sec:con}
In this work, we have investigated the influence of the quantum parameter $\xi$ which arises from the LQG on timelike geodesic motion in rotating black hole spacetimes constructed via the Newman–Janis algorithm from two different spherically symmetric LQG-inspired seed metrics. Building upon this analysis, we further examined how $\xi$ affects the gravitational waveforms emitted by periodic orbits, computed within the leading order post-Newtonian approximation.

First, we established that, under the physical requirements of horizon existence and the simultaneous presence of both MBOs and ISCOs, the admissible range of $\xi$ is bounded, i.e., $\xi_{c}=\min{\left(\xi_e,\,\sqrt{12\sqrt{3}}M\right)}$ for type I and $\xi_{c}=\min(\xi_e,\, 2\sqrt{28+19\sqrt{19}}M/3)$ for type II black hole. Moreover, the allowed interval of $\xi$ shrinks as $a$ increases. This can be understood physically as follows: both the spin parameter $a$ and the regularization parameter $\xi$ deform the geometry away from Schwarzschild, but in complementary ways. The spin introduces frame-dragging and reduces the effective potential barrier near the horizon, while $\xi$ encodes the holonomy correction that smooths the central singularity. For the black hole to maintain a regular event horizon structure, the combined deformation must not be too. As $a \to M$ (extremal Kerr limit), the horizon structure is already maximally deformed by rotation, leaving no room for additional quantum correction, so $\xi \to 0$. Conversely, for $a \to 0$, the geometry is nearly spherically symmetric and can accommodate large $\xi$. For small values of $a$, a larger $\xi$ induces a more significant variation in the specific angular momentum $L$ of particles on MBOs and ISCOs, indicating that the quantum correction encoded in $\xi$ exerts a stronger influence in the low-spin regime. This trend holds qualitatively for both type I and type II black holes. However, for intermediate spin values, the dependence of $L_{MBO(ISCO)}$ on $\xi$ is comparatively weaker in type II rotating black hole spacetimes than in type I.

Then, we extended our analysis to periodic orbits, both in the equatorial plane and in generic off-equatorial configurations. In the equatorial case, for fixed $a$ and prescribed angular momentum $L$, an increase in $\xi$ consistently enlarges the permissible energy range for bound periodic motion, which again reflects the underlying sensitivity of orbital dynamics to $\xi$. Furthermore, we found that the effects of $a$ and  $\xi$ on orbital characteristics are generally opposing: increasing one tends to counteract the influence of the other. In the non-equatorial setting, for fixed $L$ and energy $E$, the allowed range of the Carter constant $\mathcal{C}$ monotonically decreases with increasing $\xi$, irrespective of the prograde, retrograde orbits or black hole types. This suggests that larger $\xi$ effectively confines particle motion closer to the equatorial plane.

Finally, as a first step, within a simplified EMRI framework based on the linear leading order PN approximation and neglecting radiative backreaction, we computed representative gravitational waveforms for small-spin orbits around both black hole types. The results corroborate our earlier findings: the waveform morphology becomes increasingly sensitive to $\xi$ as its value grows, and waveforms generated closer to the event horizon exhibit sharper features. Notably, the imprint of $\xi$ is more pronounced in type II black holes than in type I, reflecting the distinct phenomenological signatures induced by different LQG-inspired geometries. We then show the absolute frequency spectra for such gravitational waveforms and compare the characteristic strain with various space or ground based detectors. For the selected periodic orbits, the characteristic frequencies for both type black holes concentrates between $(10^{-3},\,0.1)$ Hz but the corresponding characteristic strains are beyond the range of most detectors. In order to further investigate the influence of $\xi$ on the particle orbits and to simplify the cumbersome calculations, we considered the nearly circular orbital motion of a small mass object on the equatorial plane under the adiabatic approximation. Our focus is placed on the effect of $\xi$ on the apastron $r_{a}$. The results once again confirm that the effects of $\xi$ and $a$ on the particle orbital motion are opposite. 

A particularly notable finding is the systematic opposition between the effects of $\xi$ and $a$ on orbital dynamics. This degeneracy has important implications for parameter estimation in future EMRI observations: a measurement of orbital parameters alone may not uniquely determine both $\xi$ and $a$ without additional observational constraints, such as from black hole shadow measurements or quasi-normal mode spectroscopy. Breaking this degeneracy will require multi-messenger observations combining gravitational wave data with electromagnetic counterparts.

In summary, our results demonstrate that the regularization parameter $\xi$ leaves observable imprints on both orbital structure as well as gravitational wave emission, offering a potential observational perspective into quantum-gravitational corrections near rotating black holes. However, our investigation is still limited to the simplified approximation in which the gravitational wave’s backreaction to the particle’s motion is ignored for simplicity or only the quadrupole approximation is considered. This consideration is reasonable if we only consider one complete period of the orbital motion. A more realistic treatment incorporating radiation reaction and higher-order waveform modeling is needed. Meanwhile, several important extensions of this work are warranted. On the waveform modeling side, the leading-order post-Newtonian approximation employed here should be replaced by numerical kludge or Teukolsky-based waveforms valid in the strong-field regime, where holonomy corrections are largest. The adiabatic inspiral analysis should be extended to include eccentricity evolution and generic inclined orbits, and post-adiabatic corrections to the orbital phase should be computed. A Fisher matrix or Bayesian parameter estimation study using these improved waveforms would quantify the detectability of $\xi$ and reveal the extent of its degeneracy with the spin parameter $a$, which our results suggest is a fundamental feature of holonomy-corrected rotating spacetimes. On the theoretical side, the stability of these LQG metrics under linear perturbations and their thermodynamic consistency with full LQG should be established. Beyond gravitational waves, combining EMRI constraints on $\xi$ with complementary probes - including black hole shadow measurements from the EHT and quasi-normal mode spectroscopy - would enable multi-messenger tests of loop quantum gravity in the strong-field regime. Together, these developments would substantially advance the program of testing quantum gravity with next-generation astrophysical observations.

\begin{acknowledgments}
This work is partly supported by Natural Science Foundation of China under Grants No.12375054. YZ is also supported by the research funds No. 2055072312. Yu Han is supported by key scientific research projects in universities of Henan Province (Grant No.25A140014) and Natural Science Foundation of Henan‌ Province (Grant No.262300421865) and Nanhu Scholars Program for Young Scholars of Xinyang Normal University.
\end{acknowledgments}

\appendix
    
\section{Derivation of the expression of the extremal horizon radius $r_c$ }
First, 
$h(r)$ can be written as
\begin{eqnarray}\label{eq:hr}
 h(r)&=&r^2[(r-M)^2+(a^2-M^2)]+\xi^2(r-2M)^2.\nonumber \\
\end{eqnarray}
Obviously, we must have $a< M$ in order that $h(r)$ can vanish. It is easy to see that $h(r)> 0$ for $r\in [0,M-\sqrt{M^2-a^2}]\cup[M+\sqrt{M^2-a^2},\infty]$.

The derivative of $h(r)$ with respect to $r$ reads
\begin{eqnarray}\label{eq:hpr}
 h'(r)&=&4r\left[\left(r-\frac{3}{4}M\right)^2+\left(\frac{1}{2}a^2-\frac{9}{16}M^2\right)\right]\nonumber\\
 &&+2\xi^2(r-2M).
\end{eqnarray}
It is not difficult to check that $h'(r)<0$ for $r\in [M-\sqrt{M^2-a^2},\frac{3}{4}M+\sqrt{\frac{9}{16}M^2-\frac{a^2}{2}}]$ and $h'(r)>0$ for $r\in \left[2M,\infty\right]$.

Moreover, we have
\begin{eqnarray}\label{eq:hppr}
 h''(r)&=&12\left[\left(r-\frac{M}{2}\right)^2+\left(\frac{a^2}{6}-\frac{M^2}{4}\right)\right]+2\xi^2,
\end{eqnarray}
from which we find that $h''(r)>0$ for  $r\in \left[\frac{3}{4}M+\sqrt{\frac{9}{16}M^2-\frac{a^2}{2}},\infty\right]$. 

Therefore, we find that there exists only one root of equation $h'(r)=0$  in the range $\left[\frac{3}{4}M+\sqrt{\frac{9}{16}M^2-\frac{a^2}{2}},2M\right]$. Moreover, we can also conclude that the equation $h(r)=0$ has at most two real roots in the range $r>0$.

From the equation $h(r)=0$, we find that 
\begin{eqnarray}\label{eq:xihr}
 \xi^2=-\frac{r_c^2[(r_c-M)^2+(a^2-M^2)]}{(r_c-2M)^2},
\end{eqnarray}

The extremal horizon radius $r_c$ is fixed by the following equations,
\begin{eqnarray}\label{eq:eqsofrc}
 h(r_c)=0,\,\,h'(r_c)=0,
\end{eqnarray}
from the equation $h'(r_c)=0$, we find that 
and 
\begin{equation}\label{eq:xihpr}
 \xi^2=-\frac{2r_c\left[\left(r_c-\frac{3}{4}M\right)^2+\left(\frac{1}{2}a^2-\frac{9}{16}M^2\right)\right]}{(r_c-2M)}.\nonumber \\
\end{equation}
Thus,the two equations in (\ref{eq:eqsofrc}) yield 
\begin{eqnarray}\label{eq:xihpr}
 &&\frac{r_c^2[(r_c-M)^2+(a^2-M^2)]}{(r_c-2M)^2}\nonumber\\
 =&&\frac{2r_c\left[\left(r_c-\frac{3}{4}M\right)^2+\left(\frac{1}{2}a^2-\frac{9}{16}M^2\right)\right]}{(r_c-2M)},
\end{eqnarray}
which can be simplified as
\begin{eqnarray}\label{eq:xihpr2}
r_c(r_c-2M)(r_c-3M)=2a^2M,
\end{eqnarray}

Note that Eq.(\ref{eq:xihpr2}) has three real roots.
\begin{eqnarray}\label{eq:rrscriticalr2}
   r_k=\frac{M}{3}\left(5-2\sqrt{7} \cos{\frac{\delta+(3-2k)\pi}{3}}\right),(k=0,1,2)\nonumber \\
\end{eqnarray}
where \begin{eqnarray}
\delta\equiv\arccos{\frac{27a^2-10M^2}{7\sqrt{7}M^2}}.    
\end{eqnarray}
 However, after substituting $ r_k$ into Eq. (\ref{eq:xihr}), for $r_{(0,2)}$, we find that 
\begin{eqnarray}
 -r_{(0,2)}^2\left[(r_{(0,2)}-M)^2+(a^2-M^2)\right]<0,
\end{eqnarray}
which is in conflict with the fact $\xi^2\geq 0$, therefore, $r_{(0,2)}$ must be ruled out for physical considerations, while for $r_{(1)}$, we find that in the range $a\leq 1$ we have
\begin{eqnarray}
 -r_{(1)}^2\left[(r_{(1)}-M)^2+(a^2-M^2)\right]\geq 0,
\end{eqnarray}
which yields $\xi^2\geq 0$.  Hence, $r_{(1)}\equiv r_c$ is the  unique real  root of equations in (\ref{eq:eqsofrc}). 

It is not difficult to check that when $a=M$ we get $r_c=M$ (in this case we must have $\xi=0$, which is the classical case), and  when $0<a<M$ we get $M<r_c<2M$.

Using Eq. (\ref{eq:xihr}), we obtain 
\begin{eqnarray}\label{eq:sqxic2}
 \xi^2_e=\frac{r^2_c[2r_cM-r^2_c-a^2)]}{(r_c-2M)^2}.
\end{eqnarray}
Moreover, substituting Eq. (\ref{eq:xihpr2}) into the above equation, we  can obtain 
\begin{eqnarray}
 \xi^2_e=-\frac{r^3_c(r_c-M)}{2M(r_c-2M)},
\end{eqnarray}
from which we can easily obtain the same conclusion $M<r_c<2M$.

\nocite{*}

\bibliography{ref}% Produces the bibliography via BibTeX.

@article{Zhang:2024khj,
    author = "Zhang, Cong and Lewandowski, Jerzy and Ma, Yongge and Yang, Jinsong",
    title = "{Black holes and covariance in effective quantum gravity}",
    eprint = "2407.10168",
    archivePrefix = "arXiv",
    primaryClass = "gr-qc",
    doi = "10.1103/PhysRevD.111.L081504",
    journal = "Phys. Rev. D",
    volume = "111",
    number = "8",
    pages = "L081504",
    year = "2025"
}

@article{Zhang:2024ney,
    author = "Zhang, Cong and Lewandowski, Jerzy and Ma, Yongge and Yang, Jinsong",
    title = "{Black holes and covariance in effective quantum gravity: A solution without Cauchy horizons}",
    eprint = "2412.02487",
    archivePrefix = "arXiv",
    primaryClass = "gr-qc",
    doi = "10.1103/d6ks-d576",
    journal = "Phys. Rev. D",
    volume = "112",
    number = "4",
    pages = "044054",
    year = "2025"
}

@article{Zhang:2025ccx,
    author = "Zhang, Cong and Cao, Zhoujian",
    title = "{Covariant dynamics from static spherically symmetric geometries}",
    eprint = "2506.09540",
    archivePrefix = "arXiv",
    journal="arXiv",
    primaryClass = "gr-qc",
    month = "6",
    year = "2025"
}

@article{Chen:2025aqh,
    author = "Chen, Jiawei and Yang, Jinsong",
    title = "{Periodic orbits and gravitational waveforms in quantum-corrected black hole spacetimes}",
    eprint = "2505.02660",
    archivePrefix = "arXiv",
    primaryClass = "gr-qc",
    doi = "10.1140/epjc/s10052-025-14457-7",
    journal = "Eur. Phys. J. C",
    volume = "85",
    number = "7",
    pages = "726",
    year = "2025"
}

@article{Brahma:2020eos,
    author = "Brahma, Suddhasattwa and Chen, Che-Yu and Yeom, Dong-han",
    title = "{Testing Loop Quantum Gravity from Observational Consequences of Nonsingular Rotating Black Holes}",
    eprint = "2012.08785",
    archivePrefix = "arXiv",
    primaryClass = "gr-qc",
    doi = "10.1103/PhysRevLett.126.181301",
    journal = "Phys. Rev. Lett.",
    volume = "126",
    number = "18",
    pages = "181301",
    year = "2021"
}

@book{Wald:1984rg,
    author = "Wald, Robert M.",
    title = "{General Relativity}",
    doi = "10.7208/chicago/9780226870373.001.0001",
    publisher = "Chicago Univ. Pr.",
    address = "Chicago, USA",
    year = "1984"
}

@article{Shaikh:2018kfv,
    author = "Shaikh, Rajibul",
    title = "{Shadows of rotating wormholes}",
    eprint = "1803.11422",
    archivePrefix = "arXiv",
    primaryClass = "gr-qc",
    doi = "10.1103/PhysRevD.98.024044",
    journal = "Phys. Rev. D",
    volume = "98",
    number = "2",
    pages = "024044",
    year = "2018"
}

@article{Carter:1968rr,
    author = "Carter, Brandon",
    title = "{Global structure of the Kerr family of gravitational fields}",
    doi = "10.1103/PhysRev.174.1559",
    journal = "Phys. Rev.",
    volume = "174",
    pages = "1559--1571",
    year = "1968"
}

@article{Levin:2008mq,
    author = "Levin, Janna and Perez-Giz, Gabe",
    title = "{A Periodic Table for Black Hole Orbits}",
    eprint = "0802.0459",
    archivePrefix = "arXiv",
    primaryClass = "gr-qc",
    doi = "10.1103/PhysRevD.77.103005",
    journal = "Phys. Rev. D",
    volume = "77",
    pages = "103005",
    year = "2008"
}

@article{Babak:2006uv,
    author = "Babak, Stanislav and Fang, Hua and Gair, Jonathan R. and Glampedakis, Kostas and Hughes, Scott A.",
    title = "{'Kludge' gravitational waveforms for a test-body orbiting a Kerr black hole}",
    eprint = "gr-qc/0607007",
    archivePrefix = "arXiv",
    doi = "10.1103/PhysRevD.75.024005",
    journal = "Phys. Rev. D",
    volume = "75",
    pages = "024005",
    year = "2007",
    note = "[Erratum: Phys.Rev.D 77, 04990 (2008)]"
}

@book{Poisson:2014kt,
    author = "Poisson, Eric and Will, Clifford M.",
    title = "{Gravity: Newtonian, Post-Newtonian, Relativistic}",
    isbn = "9781139507486",
    publisher="Cambridge University Press",
    year = "2014"
}

@article{Maselli:2021men,
    author = "Maselli, Andrea and Franchini, Nicola and Gualtieri, Leonardo and Sotiriou, Thomas P. and Barsanti, Susanna and Pani, Paolo",
    title = "{Detecting fundamental fields with LISA observations of gravitational waves from extreme mass-ratio inspirals}",
    eprint = "2106.11325",
    archivePrefix = "arXiv",
    primaryClass = "gr-qc",
    doi = "10.1038/s41550-021-01589-5",
    journal = "Nature Astron.",
    volume = "6",
    number = "4",
    pages = "464--470",
    year = "2022"
}

@article{Liang:2022gdk,
    author = "Liang, Dicong and Xu, Rui and Mai, Zhan-Feng and Shao, Lijing",
    title = "{Probing vector hair of black holes with extreme-mass-ratio inspirals}",
    eprint = "2212.09346",
    archivePrefix = "arXiv",
    primaryClass = "gr-qc",
    doi = "10.1103/PhysRevD.107.044053",
    journal = "Phys. Rev. D",
    volume = "107",
    number = "4",
    pages = "044053",
    year = "2023"
}

@article{Chua:2017ujo,
    author = "Chua, Alvin J. K. and Moore, Christopher J. and Gair, Jonathan R.",
    title = "{Augmented kludge waveforms for detecting extreme-mass-ratio inspirals}",
    eprint = "1705.04259",
    archivePrefix = "arXiv",
    primaryClass = "gr-qc",
    doi = "10.1103/PhysRevD.96.044005",
    journal = "Phys. Rev. D",
    volume = "96",
    number = "4",
    pages = "044005",
    year = "2017"
}

@article{LIGOScientific:2016lio,
    author = "Abbott, B. P. and others",
    collaboration = "LIGO Scientific, Virgo",
    title = "{Tests of general relativity with GW150914}",
    eprint = "1602.03841",
    archivePrefix = "arXiv",
    primaryClass = "gr-qc",
    reportNumber = "LIGO-P1500213",
    doi = "10.1103/PhysRevLett.116.221101",
    journal = "Phys. Rev. Lett.",
    volume = "116",
    number = "22",
    pages = "221101",
    year = "2016",
    note = "[Erratum: Phys.Rev.Lett. 121, 129902 (2018)]"
}

@article{LIGOScientific:2016sjg,
    author = "Abbott, B. P. and others",
    collaboration = "LIGO Scientific, Virgo",
    title = "{GW151226: Observation of Gravitational Waves from a 22-Solar-Mass Binary Black Hole Coalescence}",
    eprint = "1606.04855",
    archivePrefix = "arXiv",
    primaryClass = "gr-qc",
    reportNumber = "LIGO-P151226",
    doi = "10.1103/PhysRevLett.116.241103",
    journal = "Phys. Rev. Lett.",
    volume = "116",
    number = "24",
    pages = "241103",
    year = "2016"
}

@article{LIGOScientific:2016aoc,
    author = "Abbott, B. P. and others",
    collaboration = "LIGO Scientific, Virgo",
    title = "{Observation of Gravitational Waves from a Binary Black Hole Merger}",
    eprint = "1602.03837",
    archivePrefix = "arXiv",
    primaryClass = "gr-qc",
    reportNumber = "LIGO-P150914",
    doi = "10.1103/PhysRevLett.116.061102",
    journal = "Phys. Rev. Lett.",
    volume = "116",
    number = "6",
    pages = "061102",
    year = "2016"
}

@article{LIGOScientific:2019fpa,
    author = "Abbott, B. P. and others",
    collaboration = "LIGO Scientific, Virgo",
    title = "{Tests of General Relativity with the Binary Black Hole Signals from the LIGO-Virgo Catalog GWTC-1}",
    eprint = "1903.04467",
    archivePrefix = "arXiv",
    primaryClass = "gr-qc",
    reportNumber = "LIGO-P1800316",
    doi = "10.1103/PhysRevD.100.104036",
    journal = "Phys. Rev. D",
    volume = "100",
    number = "10",
    pages = "104036",
    year = "2019"
}

@article{LIGOScientific:2021sio,
    author = "Abbott, R. and others",
    collaboration = "LIGO Scientific, VIRGO, KAGRA",
    title = "{Tests of General Relativity with GWTC-3}",
    eprint = "2112.06861",
    archivePrefix = "arXiv",
    primaryClass = "gr-qc",
    reportNumber = "LIGO-P2100275",
    doi = "10.1103/PhysRevD.112.084080",
    journal = "Phys. Rev. D",
    volume = "112",
    number = "8",
    pages = "084080",
    year = "2025"
}

@article{LIGOScientific:2020tif,
    author = "Abbott, R. and others",
    collaboration = "LIGO Scientific, Virgo",
    title = "{Tests of general relativity with binary black holes from the second LIGO-Virgo gravitational-wave transient catalog}",
    eprint = "2010.14529",
    archivePrefix = "arXiv",
    primaryClass = "gr-qc",
    reportNumber = "LIGO-P2000091",
    doi = "10.1103/PhysRevD.103.122002",
    journal = "Phys. Rev. D",
    volume = "103",
    number = "12",
    pages = "122002",
    year = "2021"
}

@article{Yang:2024lmj,
    author = "Yang, Sen and Zhang, Yu-Peng and Zhu, Tao and Zhao, Li and Liu, Yu-Xiao",
    title = "{Gravitational waveforms from periodic orbits around a quantum-corrected black hole}",
    eprint = "2407.00283",
    archivePrefix = "arXiv",
    primaryClass = "gr-qc",
    doi = "10.1088/1475-7516/2025/01/091",
    journal = "JCAP",
    volume = "01",
    pages = "091",
    year = "2025"
}

@article{Berti:2015itd,
    author = "Berti, Emanuele and others",
    title = "{Testing General Relativity with Present and Future Astrophysical Observations}",
    eprint = "1501.07274",
    archivePrefix = "arXiv",
    primaryClass = "gr-qc",
    doi = "10.1088/0264-9381/32/24/243001",
    journal = "Class. Quant. Grav.",
    volume = "32",
    pages = "243001",
    year = "2015"
}

@article{Ahmed:2025shr,
    author = "Ahmed, Fazlay and Wu, Qiang and Ghosh, Sushant G. and Zhu, Tao",
    title = "{Signatures of Quantum-Corrected Black Holes in Gravitational Waves from Periodic Orbits}",
    eprint = "2512.24036",
    journal="arXiv preprint",
    archivePrefix = "arXiv",
    primaryClass = "gr-qc",
    month = "12",
    year = "2025"
}

@article{Yunes:2013dva,
    author = "Yunes, Nicol{\'a}s and Siemens, Xavier",
    title = "{Gravitational-Wave Tests of General Relativity with Ground-Based Detectors and Pulsar Timing-Arrays}",
    eprint = "1304.3473",
    archivePrefix = "arXiv",
    primaryClass = "gr-qc",
    doi = "10.12942/lrr-2013-9",
    journal = "Living Rev. Rel.",
    volume = "16",
    pages = "9",
    year = "2013"
}

@article{Krishnendu:2021fga,
    author = "Krishnendu, N. V. and Ohme, Frank",
    title = "{Testing General Relativity with Gravitational Waves: An Overview}",
    eprint = "2201.05418",
    archivePrefix = "arXiv",
    primaryClass = "gr-qc",
    doi = "10.3390/universe7120497",
    journal = "Universe",
    volume = "7",
    number = "12",
    pages = "497",
    year = "2021"
}

@article{Will:1994fb,
    author = "Will, Clifford M.",
    title = "{Testing scalar - tensor gravity with gravitational wave observations of inspiraling compact binaries}",
    eprint = "gr-qc/9406022",
    archivePrefix = "arXiv",
    reportNumber = "WUGRAV-94-6",
    doi = "10.1103/PhysRevD.50.6058",
    journal = "Phys. Rev. D",
    volume = "50",
    pages = "6058--6067",
    year = "1994"
}

@article{Saijo:1996iz,
    author = "Saijo, Motoyuki and Shinkai, Hisa-aki and Maeda, Kei-ichi",
    title = "{Gravitational waves in Brans-Dicke theory : Analysis by test particles around a Kerr black hole}",
    eprint = "gr-qc/9701001",
    archivePrefix = "arXiv",
    reportNumber = "WU-AP-64-96",
    doi = "10.1103/PhysRevD.56.785",
    journal = "Phys. Rev. D",
    volume = "56",
    pages = "785--797",
    year = "1997"
}

@article{Agullo:2020hxe,
    author = "Agullo, Ivan and Cardoso, Vitor and Rio, Adri{\'a}n Del and Maggiore, Michele and Pullin, Jorge",
    title = "{Potential Gravitational Wave Signatures of Quantum Gravity}",
    eprint = "2007.13761",
    archivePrefix = "arXiv",
    primaryClass = "gr-qc",
    doi = "10.1103/PhysRevLett.126.041302",
    journal = "Phys. Rev. Lett.",
    volume = "126",
    number = "4",
    pages = "041302",
    year = "2021"
}

@article{Deppe:2025pvd,
    author = "Deppe, Nils and Heisenberg, Lavinia and Kidder, Lawrence E. and Maibach, David and Ma, Sizheng and Moxon, Jordan and Nelli, Kyle C. and Throwe, William and Vu, Nils L.",
    title = "{Signatures of quantum gravity in gravitational wave memory}",
    eprint = "2502.20584",
    archivePrefix = "arXiv",
    primaryClass = "gr-qc",
    doi = "10.1103/7c2f-975v",
    journal = "Phys. Rev. D",
    volume = "112",
    number = "2",
    pages = "024016",
    year = "2025"
}

@article{Ashtekar:2021kfp,
    author = "Ashtekar, Abhay and Bianchi, Eugenio",
    title = "{A short review of loop quantum gravity}",
    eprint = "2104.04394",
    archivePrefix = "arXiv",
    primaryClass = "gr-qc",
    doi = "10.1088/1361-6633/abed91",
    journal = "Rept. Prog. Phys.",
    volume = "84",
    number = "4",
    pages = "042001",
    year = "2021"
}

@article{Ashtekar:2011ni,
    author = "Ashtekar, Abhay and Singh, Parampreet",
    title = "{Loop Quantum Cosmology: A Status Report}",
    eprint = "1108.0893",
    archivePrefix = "arXiv",
    primaryClass = "gr-qc",
    doi = "10.1088/0264-9381/28/21/213001",
    journal = "Class. Quant. Grav.",
    volume = "28",
    pages = "213001",
    year = "2011"
}

@inproceedings{Perez:2004hj,
    author = "Perez, Alejandro",
    title = "{Introduction to loop quantum gravity and spin foams}",
    booktitle = "{2nd International Conference on Fundamental Interactions}",
    eprint = "gr-qc/0409061",
    archivePrefix = "arXiv",
    month = "9",
    year = "2004"
}

@article{Perez:2017cmj,
    author = "Perez, Alejandro",
    title = "{Black Holes in Loop Quantum Gravity}",
    eprint = "1703.09149",
    archivePrefix = "arXiv",
    primaryClass = "gr-qc",
    doi = "10.1088/1361-6633/aa7e14",
    journal = "Rept. Prog. Phys.",
    volume = "80",
    number = "12",
    pages = "126901",
    year = "2017"
}

@article{Modesto:2005zm,
    author = "Modesto, Leonardo",
    title = "{Loop quantum black hole}",
    eprint = "gr-qc/0509078",
    archivePrefix = "arXiv",
    doi = "10.1088/0264-9381/23/18/006",
    journal = "Class. Quant. Grav.",
    volume = "23",
    pages = "5587--5602",
    year = "2006"
}

@article{Zhang:2023yps,
    author = "Zhang, Xiangdong",
    title = "{Loop Quantum Black Hole}",
    eprint = "2308.10184",
    archivePrefix = "arXiv",
    primaryClass = "gr-qc",
    doi = "10.3390/universe9070313",
    journal = "Universe",
    volume = "9",
    number = "7",
    pages = "313",
    year = "2023"
}

@article{Borges:2023fub,
    author = "Borges, H. A. and Baranov, I. P. R. and Sobrinho, F. C. and Carneiro, S.",
    title = "{Remnant loop quantum black holes}",
    eprint = "2310.01560",
    archivePrefix = "arXiv",
    primaryClass = "gr-qc",
    doi = "10.1088/1361-6382/ad210c",
    journal = "Class. Quant. Grav.",
    volume = "41",
    number = "5",
    pages = "05LT01",
    year = "2024"
}

@article{Gan:2020dkb,
    author = "Gan, Wen-Cong and Santos, Nilton O. and Shu, Fu-Wen and Wang, Anzhong",
    title = "{Properties of the spherically symmetric polymer black holes}",
    eprint = "2008.09664",
    archivePrefix = "arXiv",
    primaryClass = "gr-qc",
    doi = "10.1103/PhysRevD.102.124030",
    journal = "Phys. Rev. D",
    volume = "102",
    pages = "124030",
    year = "2020"
}

@article{Ashtekar:2018lag,
    author = "Ashtekar, Abhay and Olmedo, Javier and Singh, Parampreet",
    title = "{Quantum Transfiguration of Kruskal Black Holes}",
    eprint = "1806.00648",
    archivePrefix = "arXiv",
    primaryClass = "gr-qc",
    doi = "10.1103/PhysRevLett.121.241301",
    journal = "Phys. Rev. Lett.",
    volume = "121",
    number = "24",
    pages = "241301",
    year = "2018"
}

@article{Bojowald:2018xxu,
    author = "Bojowald, Martin and Brahma, Suddhasattwa and Yeom, Dong-han",
    title = "{Effective line elements and black-hole models in canonical loop quantum gravity}",
    eprint = "1803.01119",
    archivePrefix = "arXiv",
    primaryClass = "gr-qc",
    doi = "10.1103/PhysRevD.98.046015",
    journal = "Phys. Rev. D",
    volume = "98",
    number = "4",
    pages = "046015",
    year = "2018"
}

@article{Alesci:2019pbs,
    author = "Alesci, Emanuele and Bahrami, Sina and Pranzetti, Daniele",
    title = "{Quantum gravity predictions for black hole interior geometry}",
    eprint = "1904.12412",
    archivePrefix = "arXiv",
    primaryClass = "gr-qc",
    doi = "10.1016/j.physletb.2019.134908",
    journal = "Phys. Lett. B",
    volume = "797",
    pages = "134908",
    year = "2019"
}

@article{Assanioussi:2019twp,
    author = "Assanioussi, Mehdi and Dapor, Andrea and Liegener, Klaus",
    title = "{Perspectives on the dynamics in a loop quantum gravity effective description of black hole interiors}",
    eprint = "1908.05756",
    archivePrefix = "arXiv",
    primaryClass = "gr-qc",
    doi = "10.1103/PhysRevD.101.026002",
    journal = "Phys. Rev. D",
    volume = "101",
    number = "2",
    pages = "026002",
    year = "2020"
}

@article{Han:2023wxg,
    author = "Han, Muxin and Rovelli, Carlo and Soltani, Farshid",
    title = "{Geometry of the black-to-white hole transition within a single asymptotic region}",
    eprint = "2302.03872",
    archivePrefix = "arXiv",
    primaryClass = "gr-qc",
    doi = "10.1103/PhysRevD.107.064011",
    journal = "Phys. Rev. D",
    volume = "107",
    number = "6",
    pages = "064011",
    year = "2023"
}

@article{Frodden:2012en,
    author = {Frodden, Ernesto and Perez, Alejandro and Pranzetti, Daniele and R{\"o}ken, Christian},
    title = "{Modelling black holes with angular momentum in loop quantum gravity}",
    eprint = "1212.5166",
    archivePrefix = "arXiv",
    primaryClass = "gr-qc",
    doi = "10.1007/s10714-014-1828-6",
    journal = "Gen. Rel. Grav.",
    volume = "46",
    number = "12",
    pages = "1828",
    year = "2014"
}

@article{Gambini:2020fnd,
    author = "Gambini, Rodolfo and Mato, Esteban and Pullin, Jorge",
    title = "{Axisymmetric gravity in real Ashtekar variables: the quantum theory}",
    eprint = "2001.02698",
    archivePrefix = "arXiv",
    primaryClass = "gr-qc",
    reportNumber = "LSU-REL-010820",
    doi = "10.1088/1361-6382/ab7966",
    journal = "Class. Quant. Grav.",
    volume = "37",
    number = "11",
    pages = "115010",
    year = "2020"
}

@article{Gambini:2018ucf,
    author = "Gambini, Rodolfo and Mato, Esteban and Olmedo, Javier and Pullin, Jorge",
    title = "{Classical axisymmetric gravity in real Ashtekar variables}",
    eprint = "1812.05403",
    archivePrefix = "arXiv",
    primaryClass = "gr-qc",
    reportNumber = "LSU-REL-121318",
    doi = "10.1088/1361-6382/ab1d82",
    journal = "Class. Quant. Grav.",
    volume = "36",
    number = "12",
    pages = "125009",
    year = "2019"
}

@article{Li:2024ctu,
    author = "Li, Guo-Ping and Zheng, He-Bin and He, Ke-Jian and Jiang, Qing-Quan",
    title = "{The shadow and observational images of the non-singular rotating black holes in loop quantum gravity}",
    eprint = "2410.17295",
    archivePrefix = "arXiv",
    primaryClass = "gr-qc",
    doi = "10.1140/epjc/s10052-025-13997-2",
    journal = "Eur. Phys. J. C",
    volume = "85",
    number = "3",
    pages = "249",
    year = "2025"
}

@article{Jiang:2023img,
    author = "Jiang, Hong-Xuan and Liu, Cheng and Dihingia, Indu K. and Mizuno, Yosuke and Xu, Haiguang and Zhu, Tao and Wu, Qiang",
    title = "{Shadows of loop quantum black holes: semi-analytical simulations of loop quantum gravity effects on Sagittarius~A* and M87*}",
    eprint = "2312.04288",
    archivePrefix = "arXiv",
    primaryClass = "gr-qc",
    doi = "10.1088/1475-7516/2024/01/059",
    journal = "JCAP",
    volume = "01",
    pages = "059",
    year = "2024"
}

@article{Liu:2020ola,
    author = {Liu, Cheng and Zhu, Tao and Wu, Qiang and Jusufi, Kimet and Jamil, Mubasher and Azreg-A{\"\i}nou, Mustapha and Wang, Anzhong},
    title = "{Shadow and quasinormal modes of a rotating loop quantum black hole}",
    eprint = "2003.00477",
    archivePrefix = "arXiv",
    primaryClass = "gr-qc",
    doi = "10.1103/PhysRevD.101.084001",
    journal = "Phys. Rev. D",
    volume = "101",
    number = "8",
    pages = "084001",
    year = "2020",
    note = "[Erratum: Phys.Rev.D 103, 089902 (2021)]"
}

@article{Sekhmani:2025bsi,
    author = "Sekhmani, Yassine and Ali, Heena and Ghosh, Sushant G. and Boshkayev, Kuantay",
    title = "{Rotating charged nonsingular black holes in loop quantum gravity and their observational imprints from EHT}",
    doi = "10.1016/j.jheap.2025.100425",
    journal = "JHEAp",
    volume = "49",
    pages = "100425",
    year = "2026"
}

@article{Ali:2024ssf,
    author = "Ali, Heena and Islam, Shafqat Ul and Ghosh, Sushant G.",
    title = "{Shadows and parameter estimation of rotating quantum corrected black holes and constraints from EHT observation of M87* and Sgr A*}",
    eprint = "2410.09198",
    archivePrefix = "arXiv",
    primaryClass = "gr-qc",
    doi = "10.1016/j.jheap.2025.100367",
    journal = "JHEAp",
    volume = "47",
    pages = "100367",
    year = "2025"
}

@article{Mustafa:2025mkc,
    author = "Mustafa, G. and Ghosh, Sushant G. and Donmez, Orhan and Maurya, S. K. and Orzuev, Shakhzod and Atamurotov, Farruh",
    title = "{Testing quantum-corrected black holes with QPOs observations: a study of particle dynamics and accretion~flow}",
    eprint = "2506.16405",
    archivePrefix = "arXiv",
    primaryClass = "gr-qc",
    doi = "10.1088/1475-7516/2025/10/068",
    journal = "JCAP",
    volume = "10",
    pages = "068",
    year = "2025"
}

@article{Fu:2023drp,
    author = "Fu, Guoyang and Zhang, Dan and Liu, Peng and Kuang, Xiao-Mei and Wu, Jian-Pin",
    title = "{Peculiar properties in quasinormal spectra from loop quantum gravity effect}",
    eprint = "2301.08421",
    archivePrefix = "arXiv",
    primaryClass = "gr-qc",
    doi = "10.1103/PhysRevD.109.026010",
    journal = "Phys. Rev. D",
    volume = "109",
    number = "2",
    pages = "026010",
    year = "2024"
}

@article{Moreira:2023cxy,
    author = "Moreira, Zeus S. and Lima Junior, Haroldo C. D. and Crispino, Lu{\'\i}s C. B. and Herdeiro, Carlos A. R.",
    title = "{Quasinormal modes of a holonomy corrected Schwarzschild black hole}",
    eprint = "2302.14722",
    archivePrefix = "arXiv",
    primaryClass = "gr-qc",
    doi = "10.1103/PhysRevD.107.104016",
    journal = "Phys. Rev. D",
    volume = "107",
    number = "10",
    pages = "104016",
    year = "2023"
}

@article{Bolokhov:2023bwm,
    author = "Bolokhov, S. V.",
    title = "{Long-lived quasinormal modes and overtones{\textquoteright} behavior of holonomy-corrected black holes}",
    eprint = "2311.05503",
    archivePrefix = "arXiv",
    primaryClass = "gr-qc",
    doi = "10.1103/PhysRevD.110.024010",
    journal = "Phys. Rev. D",
    volume = "110",
    number = "2",
    pages = "024010",
    year = "2024"
}

@article{Gingrich:2024tuf,
    author = "Gingrich, Douglas M.",
    title = "{Quasinormal modes of a nonsingular spherically symmetric black hole effective model with holonomy corrections}",
    eprint = "2404.04447",
    archivePrefix = "arXiv",
    primaryClass = "gr-qc",
    doi = "10.1103/PhysRevD.110.084045",
    journal = "Phys. Rev. D",
    volume = "110",
    number = "8",
    pages = "084045",
    year = "2024"
}

@article{Yang:2024ofe,
    author = "Yang, Sen and Guo, Wen-Di and Tan, Qin and Zhao, Li and Liu, Yu-Xiao",
    title = "{Parametrized quasinormal frequencies and Hawking radiation for axial gravitational perturbations of a holonomy-corrected black hole}",
    eprint = "2406.15711",
    archivePrefix = "arXiv",
    primaryClass = "gr-qc",
    doi = "10.1103/PhysRevD.110.064051",
    journal = "Phys. Rev. D",
    volume = "110",
    number = "6",
    pages = "064051",
    year = "2024"
}

@article{Liu:2024wal,
    author = "Liu, Hao and Lai, Meng-Yun and Pan, Xiao-Yin and Huang, Hyat and Zou, De-Cheng",
    title = "{Gravitational lensing effect of black holes in effective quantum gravity}",
    eprint = "2408.11603",
    archivePrefix = "arXiv",
    primaryClass = "gr-qc",
    doi = "10.1103/PhysRevD.110.104039",
    journal = "Phys. Rev. D",
    volume = "110",
    number = "10",
    pages = "104039",
    year = "2024"
}

@article{Zhao:2024elr,
    author = "Zhao, Lai and Tang, Meirong and Xu, Zhaoyi",
    title = "{The lensing effect of quantum-corrected black hole and parameter constraints from EHT observations}",
    eprint = "2403.18606",
    archivePrefix = "arXiv",
    primaryClass = "gr-qc",
    doi = "10.1140/epjc/s10052-024-13342-z",
    journal = "Eur. Phys. J. C",
    volume = "84",
    number = "9",
    pages = "971",
    year = "2024"
}

@article{Dong:2024alq,
    author = "Dong, Yuhao",
    title = "{The gravitational lensing by rotating black holes in loop quantum gravity}",
    doi = "10.1016/j.nuclphysb.2024.116612",
    journal = "Nucl. Phys. B",
    volume = "1005",
    pages = "116612",
    year = "2024"
}

@article{Soares:2024rhp,
    author = "Soares, A. R. and Vit{\'o}ria, R. L. L. and Pereira, C. F. S.",
    title = "{Topologically charged holonomy corrected Schwarzschild black hole lensing}",
    eprint = "2408.03217",
    archivePrefix = "arXiv",
    primaryClass = "gr-qc",
    doi = "10.1103/PhysRevD.110.084004",
    journal = "Phys. Rev. D",
    volume = "110",
    number = "8",
    pages = "084004",
    year = "2024"
}

@article{Ahmed:2024fye,
    author = "Ahmed, Faizuddin",
    title = "{Gravitational lensing in holonomy corrected spherically symmetric black holes with phantom global monopoles}",
    eprint = "2409.05897",
    archivePrefix = "arXiv",
    primaryClass = "gr-qc",
    doi = "10.1142/S0219887824503365",
    journal = "Int. J. Geom. Meth. Mod. Phys.",
    volume = "22",
    number = "05",
    pages = "2450336",
    year = "2025"
}

@article{Sahu:2015dea,
    author = "Sahu, Satyabrata and Lochan, Kinjalk and Narasimha, D.",
    title = "{Gravitational lensing by self-dual black holes in loop quantum gravity}",
    eprint = "1502.05619",
    archivePrefix = "arXiv",
    primaryClass = "gr-qc",
    doi = "10.1103/PhysRevD.91.063001",
    journal = "Phys. Rev. D",
    volume = "91",
    pages = "063001",
    year = "2015"
}

@article{KumarWalia:2022ddq,
    author = "Kumar Walia, Rahul",
    title = "{Observational predictions of LQG motivated polymerized black holes and constraints from Sgr A* and M87*}",
    eprint = "2207.02106",
    archivePrefix = "arXiv",
    primaryClass = "gr-qc",
    doi = "10.1088/1475-7516/2023/03/029",
    journal = "JCAP",
    volume = "03",
    pages = "029",
    year = "2023"
}

@article{Kumar:2023jgh,
    author = "Kumar, Jitendra and Islam, Shafqat Ul and Ghosh, Sushant G.",
    title = "{Strong gravitational lensing by loop quantum gravity motivated rotating black holes and EHT observations}",
    eprint = "2305.04336",
    archivePrefix = "arXiv",
    primaryClass = "gr-qc",
    doi = "10.1140/epjc/s10052-023-12205-3",
    journal = "Eur. Phys. J. C",
    volume = "83",
    number = "11",
    pages = "1014",
    year = "2023"
}

@article{Fu:2021fxn,
    author = "Fu, Qi-Ming and Zhang, Xin",
    title = "{Gravitational lensing by a black hole in effective loop quantum gravity}",
    eprint = "2111.07223",
    archivePrefix = "arXiv",
    primaryClass = "gr-qc",
    doi = "10.1103/PhysRevD.105.064020",
    journal = "Phys. Rev. D",
    volume = "105",
    number = "6",
    pages = "064020",
    year = "2022"
}

@article{Junior:2023xgl,
    author = "Junior, Ednaldo L. B. and Lobo, Francisco S. N. and Rodrigues, Manuel E. and Vieira, Henrique A.",
    title = "{Gravitational lens effect of a holonomy corrected Schwarzschild black hole}",
    eprint = "2309.02658",
    archivePrefix = "arXiv",
    primaryClass = "gr-qc",
    doi = "10.1103/PhysRevD.109.024004",
    journal = "Phys. Rev. D",
    volume = "109",
    number = "2",
    pages = "024004",
    year = "2024"
}

@article{Mushtaq:2025shw,
    author = "Mushtaq, Farzan and Tiecheng, Xia and Javed, Faisal and Ditta, Allah and Almutairi, Bander and Mustafa, G. and Hakimov, Abdullo",
    title = "{Impact of loop quantum gravity on gravitational lensing, thermal fluctuations, tidal force and geodesic deviation around a black hole}",
    doi = "10.1140/epjc/s10052-025-14281-z",
    journal = "Eur. Phys. J. C",
    volume = "85",
    number = "6",
    pages = "694",
    year = "2025",
    note = "[Erratum: Eur.Phys.J.C 85, 877 (2025)]"
}

@article{Soares:2025hpy,
    author = "Soares, A. R. and Pereira, C. F. S. and Vit{\'o}ria, R. L. L. and Silva, Marcos V. de S. and Belich, H.",
    title = "{Light deflection and gravitational lensing effects inspired by loop quantum gravity}",
    eprint = "2503.06373",
    archivePrefix = "arXiv",
    primaryClass = "gr-qc",
    doi = "10.1088/1475-7516/2025/06/034",
    journal = "JCAP",
    volume = "06",
    pages = "034",
    year = "2025"
}

@article{Alloqulov:2025htt,
    author = "Alloqulov, Mirzabek and Isaqjonov, Yokubjon and Shaymatov, Sanjar and Jawad, Abdul",
    title = "{Shadow and gravitational weak lensing around a quantum-corrected black hole surrounded by plasma}",
    doi = "10.1088/1674-1137/ad9f44",
    journal = "Chin. Phys. C",
    volume = "49",
    number = "4",
    pages = "045104",
    year = "2025"
}

@article{Li:2024afr,
    author = "Li, Haida and Zhang, Xiangdong",
    title = "{Gravitational Lensing Effects from Models of Loop Quantum Gravity with Rigorous Quantum Parameters}",
    doi = "10.3390/universe10110421",
    journal = "Universe",
    volume = "10",
    number = "11",
    pages = "421",
    year = "2024"
}

@article{Jiang:2024cpe,
    author = "Jiang, Hanyu and Alloqulov, Mirzabek and Wu, Qiang and Shaymatov, Sanjar and Zhu, Tao",
    title = "{Periodic orbits and plasma effects on gravitational weak lensing by self-dual black hole in loop quantum gravity}",
    doi = "10.1016/j.dark.2024.101627",
    journal = "Phys. Dark Univ.",
    volume = "46",
    pages = "101627",
    year = "2024"
}

@article{Huang:2022iwl,
    author = "Huang, Yang and Cao, Zhoujian",
    title = "{Finite-distance gravitational deflection of massive particles by a rotating black hole in loop quantum gravity}",
    eprint = "2212.04254",
    archivePrefix = "arXiv",
    primaryClass = "gr-qc",
    doi = "10.1140/epjc/s10052-023-11180-z",
    journal = "Eur. Phys. J. C",
    volume = "83",
    number = "1",
    pages = "80",
    year = "2023"
}

@article{Huang:2026igb,
    author = "Huang, Shilong and Chen, Jiawei and Yang, Jinsong",
    title = "{Gravitational waveforms and accretion characteristics in a quantum-corrected black hole without Cauchy horizons}",
    eprint = "2603.09140",
    journal="arXiv preprint",
    archivePrefix = "arXiv",
    primaryClass = "gr-qc",
    month = "3",
    year = "2026"
}

@article{Hu:2017mde,
    author = "Hu, Wen-Rui and Wu, Yue-Liang",
    title = "{The Taiji Program in Space for gravitational wave physics and the nature of gravity}",
    doi = "10.1093/nsr/nwx116",
    journal = "Natl. Sci. Rev.",
    volume = "4",
    number = "5",
    pages = "685--686",
    year = "2017"
}

@article{TianQin:2015yph,
    author = "Luo, Jun and others",
    collaboration = "TianQin",
    title = "{TianQin: a space-borne gravitational wave detector}",
    eprint = "1512.02076",
    archivePrefix = "arXiv",
    primaryClass = "astro-ph.IM",
    doi = "10.1088/0264-9381/33/3/035010",
    journal = "Class. Quant. Grav.",
    volume = "33",
    number = "3",
    pages = "035010",
    year = "2016"
}

@article{Robson:2018ifk,
    author = "Robson, Travis and Cornish, Neil J. and Liu, Chang",
    title = "{The construction and use of LISA sensitivity curves}",
    eprint = "1803.01944",
    archivePrefix = "arXiv",
    primaryClass = "astro-ph.HE",
    doi = "10.1088/1361-6382/ab1101",
    journal = "Class. Quant. Grav.",
    volume = "36",
    number = "10",
    pages = "105011",
    year = "2019"
}

@article{Torres-Orjuela:2023hfd,
    author = "Torres-Orjuela, Alejandro and Huang, Shun-Jia and Liang, Zheng-Cheng and Liu, Shuai and Wang, Hai-Tian and Ye, Chang-Qing and Hu, Yi-Ming and Mei, Jianwei",
    title = "{Detection of astrophysical gravitational wave sources by TianQin and LISA}",
    eprint = "2307.16628",
    archivePrefix = "arXiv",
    primaryClass = "gr-qc",
    doi = "10.1007/s11433-023-2308-x",
    journal = "Sci. China Phys. Mech. Astron.",
    volume = "67",
    number = "5",
    pages = "259511",
    year = "2024"
}

@article{Gong:2021gvw,
    author = "Gong, Yungui and Luo, Jun and Wang, Bin",
    title = "{Concepts and status of Chinese space gravitational wave detection projects}",
    eprint = "2109.07442",
    archivePrefix = "arXiv",
    primaryClass = "astro-ph.IM",
    doi = "10.1038/s41550-021-01480-3",
    journal = "Nature Astron.",
    volume = "5",
    number = "9",
    pages = "881--889",
    year = "2021"
}

@article{Babak:2017tow,
    author = "Babak, Stanislav and Gair, Jonathan and Sesana, Alberto and Barausse, Enrico and Sopuerta, Carlos F. and Berry, Christopher P. L. and Berti, Emanuele and Amaro-Seoane, Pau and Petiteau, Antoine and Klein, Antoine",
    title = "{Science with the space-based interferometer LISA. V: Extreme mass-ratio inspirals}",
    eprint = "1703.09722",
    archivePrefix = "arXiv",
    primaryClass = "gr-qc",
    doi = "10.1103/PhysRevD.95.103012",
    journal = "Phys. Rev. D",
    volume = "95",
    number = "10",
    pages = "103012",
    year = "2017"
}

@article{Hughes:2000ssa,
    author = "Hughes, Scott A.",
    editor = "Schutz, Bernard F.",
    title = "{Gravitational waves from extreme mass ratio inspirals: Challenges in mapping the space-time of massive, compact objects}",
    eprint = "gr-qc/0008058",
    archivePrefix = "arXiv",
    doi = "10.1088/0264-9381/18/19/314",
    journal = "Class. Quant. Grav.",
    volume = "18",
    pages = "4067--4074",
    year = "2001"
}

@article{Junior:2020lya,
    author = "Junior, Haroldo C. D. Lima and Crispino, Lu{\'\i}s C. B. and Cunha, Pedro V. P. and Herdeiro, Carlos A. R.",
    title = "{Spinning black holes with a separable Hamilton{\textendash}Jacobi equation from a modified Newman{\textendash}Janis algorithm}",
    eprint = "2011.07301",
    archivePrefix = "arXiv",
    primaryClass = "gr-qc",
    doi = "10.1140/epjc/s10052-020-08572-w",
    journal = "Eur. Phys. J. C",
    volume = "80",
    number = "11",
    pages = "1036",
    year = "2020"
}

@article{Azreg-Ainou:2014pra,
    author = {Azreg-A{\"\i}nou, Mustapha},
    title = "{Generating rotating regular black hole solutions without complexification}",
    eprint = "1405.2569",
    archivePrefix = "arXiv",
    primaryClass = "gr-qc",
    doi = "10.1103/PhysRevD.90.064041",
    journal = "Phys. Rev. D",
    volume = "90",
    number = "6",
    pages = "064041",
    year = "2014"
}

@article{Chaturvedi:2023ctn,
    author = "Chaturvedi, Pankaj and Kumar, Utkarsh and Thattarampilly, Udaykrishna and Kakkat, Vishnu",
    title = "{Exact rotating black hole solutions for f(R) gravity by modified Newman Janis algorithm}",
    eprint = "2309.17044",
    archivePrefix = "arXiv",
    primaryClass = "gr-qc",
    doi = "10.1140/epjc/s10052-023-12306-z",
    journal = "Eur. Phys. J. C",
    volume = "83",
    number = "12",
    pages = "1124",
    year = "2023",
    note = "[Erratum: Eur.Phys.J.C 84, 1157 (2024)]"
}

@article{Hansen:2013owa,
    author = "Hansen, Devin and Yunes, Nicolas",
    title = "{Applicability of the Newman-Janis Algorithm to Black Hole Solutions of Modified Gravity Theories}",
    eprint = "1308.6631",
    archivePrefix = "arXiv",
    primaryClass = "gr-qc",
    doi = "10.1103/PhysRevD.88.104020",
    journal = "Phys. Rev. D",
    volume = "88",
    number = "10",
    pages = "104020",
    year = "2013"
}

@article{Erbin:2016lzq,
    author = "Erbin, Harold",
    title = "{Janis-Newman algorithm: generating rotating and NUT charged black holes}",
    eprint = "1701.00037",
    archivePrefix = "arXiv",
    primaryClass = "gr-qc",
    reportNumber = "LPTENS-17-02",
    doi = "10.3390/universe3010019",
    journal = "Universe",
    volume = "3",
    number = "1",
    pages = "19",
    year = "2017"
}

@article{Al-Badawi:2025yum,
    author = "Al-Badawi, Ahmad and Ahmed, Faizuddin and Xamidov, Tursunali and Shaymatov, Sanjar and Sakall{\i}, {\.I}zzet",
    title = "{Shadow properties and orbital dynamics around an effective quantum-modified black hole surrounded by quintessential dark energy}",
    eprint = "2503.18027",
    journal="arXiv preprint",
    archivePrefix = "arXiv",
    primaryClass = "gr-qc",
    month = "3",
    year = "2025"
}

@article{Agrawal:2026rwu,
    author = "Agrawal, Rishav and Kar, Anjan and Jana, Soumya and Kar, Sayan",
    title = "{Gravitational wave radiation from periodic orbits in regular black holes}",
    eprint = "2602.20745",
    journal="arXiv preprint",
    archivePrefix = "arXiv",
    primaryClass = "gr-qc",
    month = "2",
    year = "2026"
}

@article{Chen:2026kbn,
    author = "Chen, Ruo-Ting and Fu, Guoyang and Zhang, Dan and Wu, Jian-Pin",
    title = "{Imprints of quantum gravity effects on gravitational waves: a comparative study using extreme mass-ratio inspirals}",
    eprint = "2601.00185",
    archivePrefix = "arXiv",
    journal="arXiv preprint",
    primaryClass = "gr-qc",
    month = "1",
    year = "2026"
}

@article{Huang:2025gia,
    author = "Huang, Shilong and Chen, Jiawei and Yang, Jinsong",
    title = "{Image of a quantum-corrected black hole without Cauchy horizons illuminated by a static thin accretion disk}",
    eprint = "2510.09956",
    journal="arXiv preprint",
    archivePrefix = "arXiv",
    primaryClass = "gr-qc",
    month = "10",
    year = "2025"
}

@article{Chen:2025baz,
    author = "Chen, Jiawei and Yang, Jinsong",
    title = "{Motion of spinning particles around a quantum-corrected black hole without Cauchy horizons}",
    eprint = "2509.07682",
    journal="arXiv preprint",
    archivePrefix = "arXiv",
    primaryClass = "gr-qc",
    month = "9",
    year = "2025"
}

@article{Alloqulov:2025bxh,
    author = "Alloqulov, Mirzabek and Shaymatov, Sanjar and Ahmedov, Bobomurat and Zhu, Tao",
    title = "{Regular black hole{\textquoteright}s impact on the gravitational waveforms from periodic orbits}",
    eprint = "2508.05245",
    archivePrefix = "arXiv",
    primaryClass = "gr-qc",
    doi = "10.1140/epjc/s10052-025-15251-1",
    journal = "Eur. Phys. J. C",
    volume = "86",
    number = "2",
    pages = "117",
    year = "2026"
}

@article{Guo:2025scs,
    author = "Guo, Zhiyang and Lan, Chen and Liu, Yan",
    title = "{Quantum corrected geodesic motion in polymer Kerr-like spacetime}",
    eprint = "2505.00437",
    archivePrefix = "arXiv",
    primaryClass = "gr-qc",
    doi = "10.1140/epjc/s10052-025-14872-w",
    journal = "Eur. Phys. J. C",
    volume = "85",
    number = "10",
    pages = "1200",
    year = "2025"
}

@article{Vagnozzi:2022moj,
    author = "Vagnozzi, Sunny and others",
    title = "{Horizon-scale tests of gravity theories and fundamental physics from the Event Horizon Telescope image of Sagittarius A}",
    eprint = "2205.07787",
    archivePrefix = "arXiv",
    primaryClass = "gr-qc",
    reportNumber = "UCI-HEP-TR-2022-07",
    doi = "10.1088/1361-6382/acd97b",
    journal = "Class. Quant. Grav.",
    volume = "40",
    number = "16",
    pages = "165007",
    year = "2023"
}

@article{Afrin:2022ztr,
    author = "Afrin, Misba and Vagnozzi, Sunny and Ghosh, Sushant G.",
    title = "{Tests of Loop Quantum Gravity from the Event Horizon Telescope Results of Sgr A*}",
    eprint = "2209.12584",
    archivePrefix = "arXiv",
    primaryClass = "gr-qc",
    doi = "10.3847/1538-4357/acb334",
    journal = "Astrophys. J.",
    volume = "944",
    number = "2",
    pages = "149",
    year = "2023"
}

@article{Calza:2024xdh,
    author = "Calz{\`a}, Marco and Pedrotti, Davide and Vagnozzi, Sunny",
    title = "{Primordial regular black holes as all the dark matter. II. Non-time-radial-symmetric and loop quantum gravity-inspired metrics}",
    eprint = "2409.02807",
    archivePrefix = "arXiv",
    primaryClass = "gr-qc",
    doi = "10.1103/PhysRevD.111.024010",
    journal = "Phys. Rev. D",
    volume = "111",
    number = "2",
    pages = "024010",
    year = "2025"
}

@article{Wu:2025ccc,
    author = "Wu, Meng-He and Guo, Hong and Kuang, Xiao-Mei",
    title = "{Parameter constraints on Horndeski rotating black hole through quasiperiodic oscillations}",
    eprint = "2508.13974",
    archivePrefix = "arXiv",
    primaryClass = "gr-qc",
    doi = "10.1140/epjc/s10052-025-15244-0",
    journal = "Eur. Phys. J. C",
    volume = "86",
    number = "1",
    pages = "79",
    year = "2026"
}

@article{Li:2025sfe,
    author = "Li, Yong-Zhuang and Kuang, Xiao-Mei",
    title = "{The bound orbits and gravitational waveforms of timelike particles around renormalization group improved Kerr black holes}",
    eprint = "2509.07333",
    archivePrefix = "arXiv",
    primaryClass = "gr-qc",
    doi = "10.1140/epjc/s10052-026-15510-9",
    journal = "Eur. Phys. J. C",
    volume = "86",
    number = "3",
    pages = "261",
    year = "2026"
}

@article{Li:2024tld,
    author = "Li, Yong-Zhuang and Kuang, Xiao-Mei and Sang, Yu",
    title = "{Precessing and periodic timelike orbits and their potential applications in Einsteinian cubic gravity}",
    eprint = "2401.16071",
    archivePrefix = "arXiv",
    primaryClass = "gr-qc",
    doi = "10.1140/epjc/s10052-024-12895-3",
    journal = "Eur. Phys. J. C",
    volume = "84",
    number = "5",
    pages = "529",
    year = "2024"
}

@article{Wang:2025fmz,
    author = "Wang, Zi-Liang and Battista, Emmanuele",
    title = "{Dynamical features and shadows of quantum Schwarzschild black hole in effective field theories of gravity}",
    eprint = "2501.14516",
    archivePrefix = "arXiv",
    primaryClass = "gr-qc",
    doi = "10.1140/epjc/s10052-025-13833-7",
    journal = "Eur. Phys. J. C",
    volume = "85",
    number = "3",
    pages = "304",
    year = "2025"
}

@article{Ni:2024acg,
    author = "Ni, Wei-Tou",
    title = "{Space gravitational wave detection: Progress and outlook}",
    eprint = "2409.00927",
    archivePrefix = "arXiv",
    primaryClass = "gr-qc",
    doi = "10.1360/SSPMA-2024-0186",
    journal = "Sci. Sin. Phys. Mech. Astro.",
    volume = "54",
    number = "7",
    pages = "270402",
    year = "2024"
}

@article{Barack:2018yly,
    author = "Barack, Leor and others",
    title = "{Black holes, gravitational waves and fundamental physics: a roadmap}",
    eprint = "1806.05195",
    archivePrefix = "arXiv",
    primaryClass = "gr-qc",
    doi = "10.1088/1361-6382/ab0587",
    journal = "Class. Quant. Grav.",
    volume = "36",
    number = "14",
    pages = "143001",
    year = "2019"
}

@book{Creighton:2011zz,
    author = "Creighton, Jolien D. E. and Anderson, Warren G.",
    title = "{Gravitational-wave physics and astronomy: An introduction to theory, experiment and data analysis}",
    publisher="Wiley-VCH, Berlin",
    year = "2011"
}

@article{Yunes:2025xwp,
    author = "Yunes, Nicol{\'a}s and Siemens, Xavier and Yagi, Kent",
    title = "{Gravitational-wave tests of general relativity with ground-based detectors and pulsar-timing arrays}",
    doi = "10.1007/s41114-024-00054-9",
    journal = "Living Rev. Rel.",
    volume = "28",
    number = "1",
    pages = "3",
    year = "2025"
}

@article{LIGOScientific:2014pky,
    author = "Aasi, J. and others",
    collaboration = "LIGO Scientific",
    title = "{Advanced LIGO}",
    eprint = "1411.4547",
    archivePrefix = "arXiv",
    primaryClass = "gr-qc",
    doi = "10.1088/0264-9381/32/7/074001",
    journal = "Class. Quant. Grav.",
    volume = "32",
    pages = "074001",
    year = "2015"
}

@article{LIGOScientific:2007fwp,
    author = "Abbott, B. P. and others",
    collaboration = "LIGO Scientific",
    title = "{LIGO: The Laser interferometer gravitational-wave observatory}",
    eprint = "0711.3041",
    archivePrefix = "arXiv",
    primaryClass = "gr-qc",
    reportNumber = "LIGO-P070082-04",
    doi = "10.1088/0034-4885/72/7/076901",
    journal = "Rept. Prog. Phys.",
    volume = "72",
    pages = "076901",
    year = "2009"
}

@article{Amaro-Seoane:2012vvq,
    author = "Amaro-Seoane, Pau and others",
    editor = "Hannam, Mark and Sutton, Patrick and Hild, Stefan and van den Broeck, Chris",
    title = "{Low-frequency gravitational-wave science with eLISA/NGO}",
    eprint = "1202.0839",
    archivePrefix = "arXiv",
    primaryClass = "gr-qc",
    doi = "10.1088/0264-9381/29/12/124016",
    journal = "Class. Quant. Grav.",
    volume = "29",
    pages = "124016",
    year = "2012"
}

@article{Li:2024rnk,
    author = "Li, En-Kun and others",
    title = "{Gravitational wave astronomy with TianQin}",
    eprint = "2409.19665",
    archivePrefix = "arXiv",
    primaryClass = "astro-ph.GA",
    doi = "10.1088/1361-6633/adc9be",
    journal = "Rept. Prog. Phys.",
    volume = "88",
    number = "5",
    pages = "056901",
    year = "2025"
}

@article{LISA:2022kgy,
    author = "Arun, K. G. and others",
    collaboration = "LISA",
    title = "{New horizons for fundamental physics with LISA}",
    eprint = "2205.01597",
    archivePrefix = "arXiv",
    primaryClass = "gr-qc",
    doi = "10.1007/s41114-022-00036-9",
    journal = "Living Rev. Rel.",
    volume = "25",
    number = "1",
    pages = "4",
    year = "2022"
}

@article{Ni:2012eh,
    author = "Ni, Wei-Tou",
    title = "{ASTROD-GW: Overview and Progress}",
    eprint = "1212.2816",
    archivePrefix = "arXiv",
    primaryClass = "astro-ph.IM",
    doi = "10.1142/S0218271813410046",
    journal = "Int. J. Mod. Phys. D",
    volume = "22",
    pages = "1341004",
    year = "2013"
}

@article{Ishikawa:2020hlo,
    author = "Ishikawa, Tomohiro and others",
    title = "{Improvement of the target sensitivity in DECIGO by optimizing its parameters for quantum noise including the effect of diffraction loss}",
    eprint = "2012.11859",
    archivePrefix = "arXiv",
    primaryClass = "gr-qc",
    doi = "10.3390/galaxies9010014",
    journal = "Galaxies",
    volume = "9",
    number = "1",
    pages = "14",
    year = "2021"
}

@article{Luo:2019zal,
    author = "Luo, Ziren and Guo, ZongKuan and Jin, Gang and Wu, Yueliang and Hu, Wenrui",
    title = "{A brief analysis to Taiji: Science and technology}",
    doi = "10.1016/j.rinp.2019.102918",
    journal = "Results Phys.",
    volume = "16",
    pages = "102918",
    year = "2020"
}

@article{Liu:2023qap,
    author = "Liu, Chang and Ruan, Wen-Hong and Guo, Zong-Kuan",
    title = "{Confusion noise from Galactic binaries for Taiji}",
    eprint = "2301.02821",
    archivePrefix = "arXiv",
    primaryClass = "astro-ph.IM",
    doi = "10.1103/PhysRevD.107.064021",
    journal = "Phys. Rev. D",
    volume = "107",
    number = "6",
    pages = "064021",
    year = "2023"
}

@article{Essick:2025zed,
    author = "Essick, Reed and others",
    title = "{Compact binary coalescence sensitivity estimates with injection campaigns during the LIGO-Virgo-KAGRA Collaborations{\textquoteright} fourth observing run}",
    eprint = "2508.10638",
    archivePrefix = "arXiv",
    primaryClass = "gr-qc",
    doi = "10.1103/44x3-hv3y",
    journal = "Phys. Rev. D",
    volume = "112",
    number = "10",
    pages = "102001",
    year = "2025"
}

@article{LIGOScientific:2025slb,
    author = "Abac, A. G. and others",
    collaboration = "LIGO Scientific, VIRGO, KAGRA",
    title = "{GWTC-4.0: Updating the Gravitational-Wave Transient Catalog with Observations from the First Part of the Fourth LIGO-Virgo-KAGRA Observing Run}",
    eprint = "2508.18082",
    journal="arXiv",
    archivePrefix = "arXiv",
    primaryClass = "gr-qc",
    reportNumber = "LIGO-P2400386",
    month = "8",
    year = "2025"
}

@article{Peters:1964zz,
    author = "Peters, P. C.",
    title = "{Gravitational Radiation and the Motion of Two Point Masses}",
    doi = "10.1103/PhysRev.136.B1224",
    journal = "Phys. Rev.",
    volume = "136",
    pages = "B1224--B1232",
    year = "1964"
}

@article{Peters:1963ux,
    author = "Peters, P. C. and Mathews, J.",
    title = "{Gravitational radiation from point masses in a Keplerian orbit}",
    doi = "10.1103/PhysRev.131.435",
    journal = "Phys. Rev.",
    volume = "131",
    pages = "435--439",
    year = "1963"
}

@article{Zhao:2025sck,
    author = "Zhao, Lai and Tang, Meirong and Xu, Zhaoyi",
    title = "{Constraints on the scale parameter of regular black hole in asymptotically safe gravity from extreme mass ratio inspirals}",
    eprint = "2503.06503",
    archivePrefix = "arXiv",
    primaryClass = "gr-qc",
    doi = "10.1088/1475-7516/2025/10/002",
    journal = "JCAP",
    volume = "10",
    pages = "002",
    year = "2025"
}

@article{Ryan:1995zm,
    author = "Ryan, Fintan D.",
    title = "{Effect of gravitational radiation reaction on circular orbits around a spinning black hole}",
    eprint = "gr-qc/9506023",
    archivePrefix = "arXiv",
    doi = "10.1103/PhysRevD.52.R3159",
    journal = "Phys. Rev. D",
    volume = "52",
    pages = "R3159--R3162",
    year = "1995"
}

@book{Maggiore:2007ulw,
    author = "Maggiore, Michele",
    title = "{Gravitational Waves. Vol. 1: Theory and Experiments}",
    doi = "10.1093/acprof:oso/9780198570745.001.0001",
    isbn = "978-0-19-171766-6, 978-0-19-852074-0",
    publisher = "Oxford University Press",
    year = "2007"
}

@article{Liu:2024qci,
    author = "Liu, Yunlong and Zhang, Xiangdong",
    title = "{Gravitational waves for eccentric extreme mass ratio inspirals of self-dual spacetime}",
    eprint = "2404.08454",
    archivePrefix = "arXiv",
    primaryClass = "gr-qc",
    doi = "10.1088/1475-7516/2024/10/056",
    journal = "JCAP",
    volume = "10",
    pages = "056",
    year = "2024"
}

@article{Muguruza:2026hqn,
    author = "Muguruza, Pablo F. and Sopuerta, Carlos F.",
    title = "{Probing Kerr Symmetry Breaking with LISA Extreme-Mass-Ratio Inspirals}",
    eprint = "2604.06053",
    archivePrefix = "arXiv",
    journal="arXiv",
    primaryClass = "gr-qc",
    month = "4",
    year = "2026"
}

@article{Gong:2025mne,
    author = "Gong, Huajie and Long, Sheng and Wang, Xi-Jing and Xia, Zhongwu and Wu, Jian-Pin and Pan, Qiyuan",
    title = "{Gravitational waveforms from periodic orbits around a novel regular black hole}",
    eprint = "2509.23318",
    journal="arXiv",
    archivePrefix = "arXiv",
    primaryClass = "gr-qc",
    month = "9",
    year = "2025"
}

\end{document}